\documentclass[aps, prb, twocolumn, superscriptaddress, floatfix, notitlepage, a4paper, 10pt]{revtex4-2}
\usepackage[utf8]{inputenc}
\usepackage{amssymb}
\usepackage{gensymb}
\usepackage{amsmath}
\usepackage{xcolor}
\usepackage{graphicx}
\usepackage{pdfsync}
\usepackage{hyperref}
\hypersetup{colorlinks, linkcolor={blue}, citecolor={blue}, urlcolor={blue}}
\usepackage{bm}
\usepackage{comment}
\usepackage[normalem]{ulem}
\usepackage{mathptmx}
\usepackage{textcomp}
\begin{document}
\title{Materials and devices for fundamental quantum science and quantum technologies}
\date{\today}
\author{Marco Polini}
\email{marco.polini@unipi.it}
\affiliation{Dipartimento di Fisica, Universit\`a di Pisa, Largo Bruno Pontecorvo 3, I-56127 Pisa,~Italy}
\affiliation{Istituto Italiano di Tecnologia, Graphene Labs, Via Morego 30, I-16163 Genova,~Italy}
\author{Francesco Giazotto}
\email{francesco.giazotto@sns.it}
\affiliation{NEST, Istituto Nanoscienze-CNR and Scuola Normale Superiore, I-56127 Pisa,~Italy}
\author{Kin Chung Fong}
\affiliation{Quantum Engineering and Computing Group, Raytheon BBN Technologies, Cambridge, Massachusetts 02138,~USA}
\author{Ioan M. Pop}
\affiliation{Institute for Quantum Materials and Technology,
Karlsruher Institute of Technology, 76344 Eggenstein-Leopoldshafen,~Germany}
\affiliation{Physikalisches Institut, Karlsruhe Institute of Technology, 76131 Karlsruhe,~Germany}
\author{Carsten Schuck}
\affiliation{Institute of Physics, University of M\"{u}nster, Wilhelm-Klemm-Str. 10, 48149 M\"{u}nster,~Germany}
\affiliation{Center for Nanotechnology (CeNTech), Heisenbergstr. 11, 48149 M\"{u}nster,~Germany}
\affiliation{Center for Soft Nanoscience (SoN), Busso-Peus-Str. 10, 48149 M\"{u}nster,~Germany}
\author{Tommaso Boccali}
\affiliation{INFN, Sezione di Pisa, Largo Bruno Pontecorvo 3, I-56127 Pisa,~Italy}
\author{Giovanni Signorelli}
\affiliation{INFN, Sezione di Pisa, Largo Bruno Pontecorvo 3, I-56127 Pisa,~Italy}
\author{Massimo D'Elia}
\affiliation{Dipartimento di Fisica, Universit\`a di Pisa, Largo Bruno Pontecorvo 3, I-56127 Pisa,~Italy}
\affiliation{INFN, Sezione di Pisa, Largo Bruno Pontecorvo 3, I-56127 Pisa,~Italy}
\author{Robert H. Hadfield}
\affiliation{James Watt School of Engineering, University of Glasgow, Glasgow G12 8QQ,~United Kingdom}
\author{Vittorio Giovannetti}
\affiliation{NEST, Scuola Normale Superiore and Istituto Nanoscienze-CNR, I-56127 Pisa,~Italy}
\author{Davide Rossini}
\affiliation{Dipartimento di Fisica, Universit\`a di Pisa, Largo Bruno Pontecorvo 3, I-56127 Pisa,~Italy}
\affiliation{INFN, Sezione di Pisa, Largo Bruno Pontecorvo 3, I-56127 Pisa,~Italy}
\author{Alessandro Tredicucci}
\affiliation{Dipartimento di Fisica, Universit\`a di Pisa, Largo Bruno Pontecorvo 3, I-56127 Pisa,~Italy}
\affiliation{NEST, Istituto Nanoscienze-CNR and Scuola Normale Superiore, I-56127 Pisa,~Italy}
\author{Dmitri K. Efetov}
\affiliation{ICFO-Institut de Ci\`{e}ncies Fot\`{o}niques, The Barcelona Institute of Science and Technology, Av. Carl Friedrich Gauss 3, 08860 Castelldefels (Barcelona),~Spain}
\author{Frank H. L. Koppens}
\affiliation{ICFO-Institut de Ci\`{e}ncies Fot\`{o}niques, The Barcelona Institute of Science and Technology, Av. Carl Friedrich Gauss 3, 08860 Castelldefels (Barcelona),~Spain}
\affiliation{ICREA-Instituci\'{o} Catalana de Recerca i Estudis Avan\c{c}ats, Passeig Llu\'{i}s Companys 23, 08010 Barcelona,~Spain}
\author{Pablo Jarillo-Herrero}
\affiliation{Department of Physics, Massachusetts Institute of Technology, Cambridge, Massachusetts 02139,~USA}
\author{Anna Grassellino}
\affiliation{Superconducting Quantum Materials and Systems Center, Fermi National Accelerator Laboratory, Batavia, Illinois 60510,~USA}
\author{Dario Pisignano}
\affiliation{Dipartimento di Fisica, Universit\`a di Pisa, Largo Bruno Pontecorvo 3, I-56127 Pisa,~Italy}
\affiliation{NEST, Istituto Nanoscienze-CNR and Scuola Normale Superiore, I-56127 Pisa,~Italy}

\begin{abstract}
{\bf Technologies operating on the basis of quantum mechanical laws and resources such as phase coherence and entanglement are expected to revolutionize our future. Quantum technologies are often divided into four main pillars:  computing, simulation, communication, and sensing \& metrology. Moreover, a great deal of interest is currently also nucleating around energy-related quantum technologies. In this Perspective, we focus on advanced superconducting materials, van der Waals materials, and moir\'e quantum matter, summarizing recent exciting developments and highlighting a wealth of potential applications, ranging from high-energy experimental and theoretical physics to quantum materials science and energy storage.} 
\end{abstract}

\maketitle

 Quantum information science, a very active research field stemming from the convergence of quantum mechanics and information science that developed towards the end of the twentieth century, has led us to a “second quantum revolution”~\cite{Aspect2003,Dowling2003,Deutsch2020}. The promised outcome is a portfolio of next-generation quantum technologies that can outperform current systems based on the discoveries made during the “first quantum revolution,” such as semiconductor devices and lasers. 
 
 In these present times, the acceleration of the second quantum revolution is truly tangible. We are currently witnessing an overwhelming proliferation of sensational headlines about game changing developments in quantum technology. Furthermore, quantum technology programs are heavily financed throughout the globe, from the United States~\cite{Raymer2019} to the United Kingdom~\cite{Knight2019}, Europe~\cite{Riedel2019}, and China~\cite{Zhang2019_Intro}. In the midst of this global excitement,  with scientists urged to raise barricades against marketing zingers, this Perspective has a very humble and down-to-earth scope. We simply offer an interdisciplinary view on recent materials science efforts towards the fabrication of high-quality devices for quantum computation, quantum communication, quantum sensing and metrology, and quantum simulation. One of the key elements of this Perspective is that we treat on equal footing these four pillars together with  recent progress in {\it quantum thermodynamics}, highlighting a number of topics related to the interplay between quantum mechanical resources and energy/work. We focus on materials science discoveries and related devices, which will hopefully yield scalable architectures, in the same way that the transistor, integrated circuits and magnetic hard disks have led to a revolution in computing power, portable electronics and the ability to store and handle huge amounts of data. We do not analyze  quantum coherent systems such as cold atoms~\cite{Schafer2020}, trapped ions~\cite{Monroe2021}, and Rydberg atoms~\cite{Adams2020}, for which recent Reviews and Perspective articles are already available.

\section{Superconducting cavities, qubits, and memories}
\label{sect:quantum_computing_hardware}

\begin{figure*}[t]
\centering
\includegraphics[width=0.8 \linewidth]{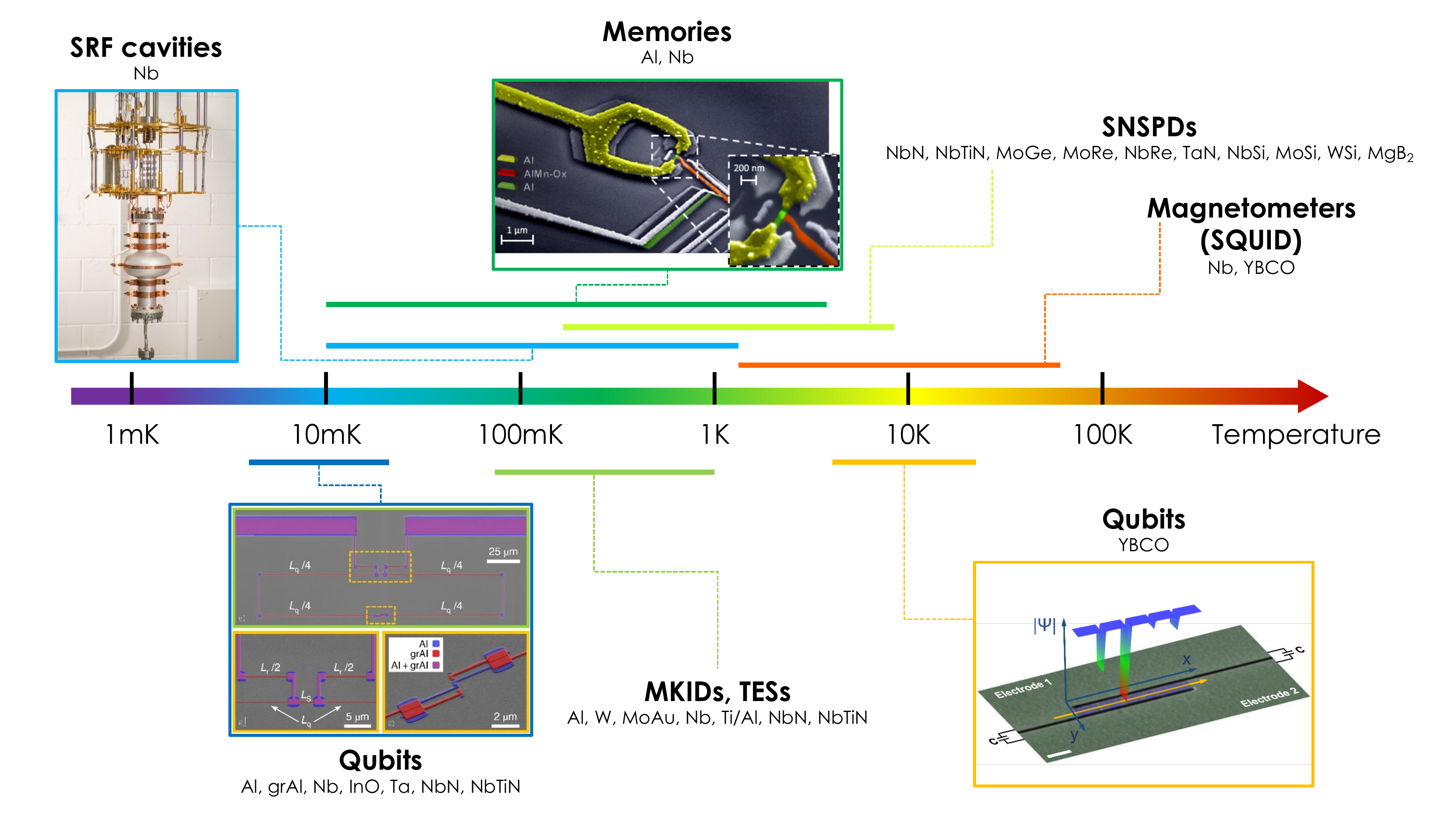}
\caption{{\bf Superconducting quantum devices for quantum technologies and their operating temperature ranges}. Top left panel: A Nb SRF cavity for, e.g., dark axion searches. Top central panel: Pseudo-color scanning electron micrograph of a prototypical persistent Josephson phase-slip memory cell (PSMC) with topological protection. The inset shows a blow-up of the PSMC in the Josephson weak-link region (green nanowire). The color legend indicates the material used for each circuit element: yellow and green for aluminum, and red for ${\rm AlMn}$-${\rm Ox}$. The PSMC has been operated from 25 mK up to about 1K. Adapted from Ref.~\cite{Ligato2021a} under the terms of the CC 4.0 license (\href{http://creativecommons.org/licenses/by/4.0}{http://creativecommons.org/licenses/by/4.0}). Bottom left panel: Scanning electron micrograph image of a fluxonium qubit built by using a granular Al (grAl) superinductor. The qubit consists of a Josephson junction shunted by a $300$-${\mu m}$-long grAl superinductor, and it is coupled through a shared inductance $L_{\rm S}$ to a microwave resonator consisting also of a grAl inductor $L_{\rm r}$ and a coplanar capacitor. The color legend indicates the material used for each circuit element: blue for aluminum, red for grAl and purple for aluminum covered with grAl. Adapted by permission from Ref.~\cite{Grunhaupt2019}. Copyright~\textcopyright~Springer~Nature~(2019). Bottom right panel (courtesy of Matvey Lyatti, University of M\"{u}nster, Germany): Scanning electron micrograph of a YBCO phase-slip nanowire (PSN) produced by focused ion beam milling. The large intrinsic capacitance C between the direct current-carrying (orange) PSN and the superconducting electrodes is required to readout the nanowire quantum state. We have also plotted typical spatial fluctuations of the order parameter $|\Psi|$ along the nanowire. Regions where $|\Psi|$ is strongly suppressed are akin to dynamical Josephson weak links. Scale bar: $500~{\rm nm}$. {\bf List of acronyms present in the figure}: SRF = Superconducting Radio Frequency;
SNSPDs = Superconducting Nanowire Single-Photon Detectors;
SQUID =  Superconducting Quantum Interference Device;
Qubits = Quantum bits;
MKIDs = Microwave Kinetic Inductance Detectors;
TESs = Transition Edge Sensors.
\label{fig:one}}
\end{figure*}

Superconducting materials are considered key candidates for a number of cutting-edge applications in quantum technology, particularly in the fields of quantum computation and radiation sensing. In this Section we summarize the most recent trends towards the implementation of high-performance qubits based on three-dimensional (3D) superconducting cavities, disordered and high-$T_{\rm c}$ superconducting films, and two-dimensional (2D) van der Waals (vdW) materials---see Fig.~\ref{fig:one}. We  then briefly address recent designs for superconducting memories, which also represent a pivotal asset for the realization of an energy-efficient superconducting computer.

\subsection{Superconducting cavities}
\label{sect:superconducting cavities}
Substantial progress has been made in the 3D circuit quantum electrodynamics (cQED) architecture~\cite{Paik2011, Reagor2016,Blais2021} whereby coupling between 3D superconducting radio frequency (SRF) cavities and 2D superconducting qubits is achieved. This architecture offers several advantages with respect to conventional 2D architectures for quantum computing and sensing. 
These include: i) A lower surface-to-volume ratio, minimizing the effect of dielectric losses; ii) A clean electromagnetic environment, the enclosing superconducting cavity serving as an effective shield; and iii) The possibility to encode several qubits inside each of the cavity-transmon modules. The latter allows to substantially decrease the required number of microwave channels for system control/manipulation. 

Recently, a large leap was achieved at Fermilab in the coherence times of 3D SRF resonators in the quantum regime, reaching lifetimes on the order of $2~{\rm s}$~\cite{Romanenko2020}. These were possible thanks to materials science advancements, in particular the abatement of two-level-system (TLS) losses in niobium via in-situ baking and dissolution of the amorphous niobium pentoxide layer. Thanks to this breakthrough, SRF cavities are now orders of magnitude more coherent than state-of-the-art superconducting qubits, creating a promising approach to build cavity-based logical qubits where superpositions of Fock states of the cavity can be engineered to encode a qubit or even a register. A simple example consists in encoding the $|0\rangle$ and $|1\rangle$ states into the presence or absence of microwave photons in specific cavity quantum states. This approach has been pioneered and successfully demonstrated by several leading groups---see e.g.~Refs.~\cite{Vlastakis2013, Ofek2016}---and could be used for error correction on qubit registers living in the 2D Fock space of an SFR cavity~\cite{Terhal2020, Cai2021}. Furthermore, the proof-of-concept demonstration of a ``random-access'' quantum processor~\cite{Naik2017} based on controlling a multi-cell 3D cavity with a single transmon opens up a promising pathway for fast scale-up via applications of multi-cell high-coherence SRF cavities as multi-qubit quantum processor units. The fact that there is plenty of room available in the quadrature space of an oscillator offers a tantalizing, possibly hardware efficient, alternative to quantum error correction compared to extended 2D qubit fabrics~\cite{Satzinger2021}. 

\subsection{Disordered superconductors for qubits}

In the last decade several disordered superconducting materials, such as NbN~\cite{Niepce2019}, NbTiN~\cite{Hazard2019}, TiN~\cite{Shearrow2018}, and granular aluminum (grAl)~\cite{Maleeva2018} have entered the scene and will probably play an important role in future devices. These materials share at least two valuable attributes for quantum circuits: i) Disorder in their structure is directly proportional to characteristic sheet inductance, enabling the realization of so-called superinductors, inductors with a characteristic impedance exceeding the resistance quantum $R_0 \equiv h/(4e^2) \simeq 6.45~{\rm kOhm}$ (where $h$ is Planck's constant and $e$ is the elementary charge), without the use of mesoscopic Josephson junctions~ \cite{Grunhaupt2019}; ii) Thanks to their relatively large London penetration depth, disordered superconductors can withstand in-plane magnetic fields, allowing them to be used in hybrid circuits to implement couplers, readout resonators, and qubits~\cite{Winkel2020, Pita-Vidal2020}. 

Due to their complex internal structure, however, disordered superconductors are in principle more exposed to dissipation and dephasing mechanisms~\cite{YangF2020}. Understanding and mitigating decoherence mechanisms in homogeneously disordered (NbN, NbTiN, TiN, InO, etc.) and grAl superconductors is a significant challenge, and it will probably be linked to the exploration of a rich landscape of quantum phenomena~\cite{Sacepe2020}.

\subsection{High-$T_{\rm c}$ superconducting qubits} 
Phase slip nanowires (PSNs), made from high-$T_{\rm c}$ materials, are promising for novel qubit implementations because they feature long-lived excited states at temperatures that exceed the operating point of most low-$T_{\rm c}$ devices by $2$-$3$ orders of magnitude. 

In nanowires made from cuprate superconductors, such as yttrium barium copper oxide (YBCO), several quantized energy levels of Josephson plasma oscillations can persist due to the fact that the superconducting energy gap is much larger than that of their low-$T_{\rm c}$ counterparts. Different from tunnel Josephson junctions, such PSNs with finite critical current are superconducting weak links with direct conductivity and nonlinear current-phase relations~\cite{Mooji2006,deGraaf2018}. 

For sufficiently small nanowire cross sections and a low density of Cooper pairs, large fluctuations of the order parameter make phase slippage the dominant mechanism for resistive states~\cite{Astafiev2012}. Small nanowire cross sections also transform the $d$-wave symmetry of the order parameter with zero-gap nodes into fully gapped states with dramatically reduced numbers of quasiparticles at low temperature, thereby yielding millisecond-long excited state lifetimes and very high quality factors~\cite{Lyatti2020}. Employing current-biased PSNs as qubits that are resilient to charge and critical current noise~\cite{Mooji2005} hence provides a perspective for a significant improvement over the coherence times of current low-temperature superconducting circuits, thus benefiting the number of executable gate operations. 

Progress with high-$T_{\rm c}$ PSNs will depend on better control over the material properties as well as improving the theoretical description of superconducting fluctuation phenomena in high-$T_{\rm c}$ systems.

\subsection{Quantum devices built by using vdW materials}  
Advancements in materials research can potentially improve qubit coherence and functionality. To date, conventional superconducting qubits are mostly fabricated by using amorphous aluminum and its oxide, which can host TLSs, limiting qubit coherence. Several possible paths have been recently followed to increase coherence:  i) Replacing the oxide~\cite{Place2021}; ii) Employing crystalline epitaxial materials such as GaAs~\cite{McRae2021} and  nitrides~\cite{Melville2020, Kim2021}; and iii) Employing 2D vdW materials~\cite{Lee19,Schmidt2018,Wang2019,Antony2021,Wang2021}.

Unlike epitaxial materials, vdW crystals are usually grown by chemical vapor deposition/transport or by the flux method~\cite{Rhodes2019}, before being assembled to a designed heterostructure by stacking. Scalable growth, transfer and wafer-scale integration have become quite mature~\cite{Akinwande2019,Asselberghs2020} in the recent years.  The vdW materials platform presents an interesting case because the weak vdW force between the layers allows for atomically sharp interfaces despite the lattice mismatch and because no chemical bonds form between the layers. Moreover, the gate tunability of vdW materials can be useful in modulating quantum devices~\cite{Schmidt2018,Wang2019}. Recently, transmons using vdW parallel-plate capacitors have demonstrated quantum coherence~\cite{Antony2021,Wang2021}. 
To avoid energy relaxation of qubits by phonon emission, non-piezoelectric crystals with inversion symmetry are preferred. Research on piezoelectric devices such as high-overtone bulk acoustic resonators (${\rm HBAR}$) and related materials such as ${\rm LiNbO}_3$ is important though for the realization of quantum transduction and memories~\cite{Gokhale2020}. 

To suppress the loss of quantum information, research for next-generation qubits will need to focus on low-loss materials, as measured by the so-called ``loss tangent'', which is inversely proportional to the maximum possible $T_1$ relaxation time of a qubit. In the microwave frequency range where superconducting qubits operate, losses increase as the probe power decreases due to mechanisms involving TLSs and quasiparticles. The maximal loss when the probe power corresponds to a single photon is most relevant to qubit relaxation because qubits operate in the single-photon regime. This aspect highlights the need for materials research under the very challenging conditions of low signal and low-noise environment. As discussed above, higher superconducting transition temperatures are also desirable because a larger superconducting gap may suppress quasiparticle generation, which is often a dominant mechanism of qubit relaxation. New concepts, such as merged-element transmons~\cite{Zhao2020} and gate tunable qubits~\cite{Wang2019}, critically demand for materials development.

\subsection{Superconducting memories}
\label{sect:superconducting_memories}

Although superconducting processors both in the quantum and classical regimes promise increased computational power, it is recognized that dense arrays of fast memory cells are one of the main limiting factors to achieve a complete superconducting computer. 

As a matter of fact, fast and scalable superconducting memories have not yet been implemented, since their operation typically relies on the inductance of their superconducting loop~\cite{Polonsky1995}, the latter being on the order of a few micrometers of diameter. In the context of fully-superconducting memories, some improvements were achieved in the last few years by exploiting either Nb-based current-controlled non-volatile kinetic inductance memory elements~\cite{Ilin2021, Murphy2017} or Al-based persistent non-volatile Josephson phase-slips memory cells~\cite{Ligato2021a}. Yet, the use of magnetic memories enables miniaturization down to the nanometer scale, but in such a case hybrid ferromagnet-superconductor elements are exploited. In particular, the operation of either Nb-based Josephson memories containing Ni~\cite{Madden2018} or PdFe~\cite{Karelina2021} ferromagnetic layers was recently demonstrated. The realization of Nb-based Josephson junctions integrated with a magnetic spin valve was also reported~\cite{Baek2014}, providing an alternative prototype for a non-volatile nanoscale memory device. Very recently, an Al-based superconducting, flux-tunable, non-volatile thermal memory cell was also reported~\cite{Ligato2021b}, which could prove its potential for applications in energy harvesting and thermal logic architectures~\cite{Fornieri2017}.

\section{Quantum computing applications}
\label{sect:quantum_computing_applications}

Scientific computing is more and more important in experimental and theoretical hard sciences, where the need to process vast amounts of data or perform complex and long {\it ab-initio} calculations increases faster than technology. Quantum Computing (QC) is a candidate solution, at least in the medium-to-long time frame.

\subsection{Computational quantum chemistry and many-body physics}

It is widely believed that QC has the potential to tackle classically intractable problems in chemistry, physics, and materials science, where electronic structure calculations play a key role~\cite{McArdle2020}. In fact, due to the increasing ambition to simulate complicated chemical compounds, chemical processes (e.g.~in electrochemical batteries) and quantum materials, and to identify molecules that exhibit significant therapeutic activity against a disease, electronic structure calculations are becoming more and more computationally expensive. The aim is to find the eigenstates and corresponding eigenvalues of an electronic Hamiltonian, including electron-electron (e-e) interactions. For example, in computational quantum chemistry one needs to deal, for a given nuclear configuration, with an electronic Hamiltonian describing electrons in a molecule, interacting with the nuclei and among each other. Solving the corresponding Schr\"{o}dinger equation for a range of nuclear configurations yields the potential energy surfaces of the molecule. Mapping out these potential energy curves explicitly is exponentially costly in the number of degrees of freedom of the molecule. In condensed matter physics and materials science, instead, one typically deals with crystals featuring Bloch translational invariance and a given number of atoms per unit cell. When the single-particle Bloch bands are narrow or flat (as in the case of the moir\'{e} quantum materials discussed below in Sect.~\ref{sect:quantum_simulation}), e-e interactions lead to strong correlations, which are difficult to tackle analytically or numerically. 

A QC approach starts by encoding first- or second-quantized fermionic Hamiltonians of chemical and physical interest onto a collection of qubits. An encoding method is a map from the fermionic Fock space to the Hilbert space of qubits such that every fermionic state can be represented by a qubit state. Several different encoding protocols are available. For example, second-quantized fermionic Hamiltonians can be mapped onto qubit Hamiltonians by using the Jordan-Wigner~\cite{Jordan1928} and Bravyi-Kitaev encodings~\cite{Bravyi2002}. After the mapping has been carried out, a quantum algorithm needs to be used for finding the ground and excited states of the chemical/physical system of interest. Examples include the variational quantum eigensolver (VQE)~\cite{Peruzzo2014} and quantum phase estimation~\cite{Kitaev1995,Abrams1999}. Finally, one needs to use suitable techniques to mitigate the effects of noise in non-error-corrected QC platforms (such as Google ``Sycamore'', IBM Quantum ``System One'', Rigetti ``Aspen'', ``Zuchongzhi'', etc), which will be crucial for achieving accurate simulations. Indeed, it is to date under debate and a topic of research whether or not such noisy intermediate-scale quantum (NISQ) devices~\cite{Preskill2018} will be able to solve classically-intractable problems of chemical and physical interest. 

Recently, Arute et al.~\cite{Arute2020} have implemented a VQE on the Google Sycamore superconducting quantum processor using up to $12$ qubits. More precisely, the experiment implements the mean-field Hartree-Fock method for calculating the binding energy of hydrogen chains and the isomerization of diazene. While it is totally feasible to solve the exact same problem on a classical computer, this work demonstrates several key building blocks for quantum chemistry simulations and paves the way for achieving quantum advantage for problems of chemical interest. We refer the reader to the excellent recent review by McArdle et al.~\cite{McArdle2020} for more details on computational quantum chemistry and an exhaustive list of references.

 In the realm of many-body physics and in view of their limited size and coherence, it seems wise in the short term to identify particular physical phenomena that NISQ devices can immediately and naturally realize, as opposed to physics they could realize universally. For example, Ippoliti et al.~\cite{Ippoliti2021} have demonstrated that the architecture of Google Sycamore is ideal to realize an elusive non-equilibrium state of matter, i.e.~the so-called ``discrete time crystal''~\cite{Else2020}. Another possible near-term application of NISQ devices is to analyze quantum critical systems~\cite{Rossini2021,Dupont2021}. The collective behavior of a quantum many-body system near a quantum phase transition (QPT) can be described, both at and out-of equilibrium conditions induced by unitary or dissipative mechanisms, via a unified {\it scaling picture} that may hold in the thermodynamic and finite-size scaling limits~\cite{Rossini2021}. For continuous QPTs such framework is based on the renormalization-group theory of critical phenomena, while in the case of first-order quantum transitions it is ruled by the avoided crossing of two levels, being characterized by an extreme sensitivity to the boundary conditions. The emerging scenario is amenable to direct laboratory testing in available NISQ devices as~e.g.~finite-size scaling emerges in small-sized systems, with a few tens of lattice sites~\cite{Rossini2021}. To mention a concrete example, the critical properties of the one-dimensional (1D) quantum Ising model have been probed on a Rigetti (Aspen-9) superconducting quantum processor via a Kibble-Zurek process (a dynamically-driven phenomenon to access quantum criticality), obtaining scaling laws and estimating critical exponents despite inherent sources of errors on the hardware~\cite{Dupont2021}.

\subsection{Computational high-energy physics}

The Standard Model of particle physics is based on quantum gauge theories. 
Its predictions affect a wide range of phenomena, ranging from the physics of the early universe to hadron phenomenology. In some cases, like e.g.~in the case of low-energy quantum chromodynamics (QCD), they can only be obtained by numerical means. 
Lattice QCD simulations have been representing a major challenge for high performance computing in the past few decades. They usually proceed through a Monte Carlo sampling of the QCD path integral, a task which is feasible on a classical computer, even if requiring resources close to the exascale. However, in some conditions, like for QCD at finite baryon	density or for non-equilibrium phenomena, the approach fails, because the path integral measure becomes complex and Monte-Carlo methods are not applicable. This is the (in)famous ``sign problem'', which has similar counterparts in condensed matter physics~\cite{Troyer2005}, and strongly limits our present knowledge, for instance about neutron star physics.

In this respect, a quantum computer would represent the ideal solution to the problem~\cite{Banuls2020}, as originally envisioned by Richard Feynman. Indeed, the sign problem is not an intrinsic problem of the physical system by itself, but just a wrong but compelling choice of the computational basis on a classical computer. A quantum computer would ideally be able to explore the space of quantum states of the system, e.g.~through a quantum Markov chain~\cite{Kemme2011, Clemente2020} or by other suitable algorithms~\cite{Yung2012, Motta2020}, so as to provide us with the information we are presently missing. 

In perspective, realistic applications would need to digitalize a 3D lattice with at least ten sites in each direction, with fermion fields living on each site and an ${\rm SU}(3)$ gauge link variable for each site and direction. The number of required qubits, considering also error correction issues, largely exceeds those that can be realistically predicted for real quantum machines in the next few years. 
The road for algorithmic and technological development is however open now.

\subsection{Data analysis in high-energy physics experiments}
In the last three decades, high-energy physics (HEP) experimental and theoretical sectors have increasingly become more computing hungry, with needs that have largely surpassed a ``constant cost envelope'', even in presence of very effective technological improvements from the vendors~\cite{Boccali2019}. The next big jump in resource needs is expected to happen when the High Luminosity Large Hadron Collider (LHC)~\cite{Schmidt2016} will enter operations, around 2028. Current estimates indicate that the cost of computing will be up to three times the amount of money spent yearly at the present~\cite{CMS}. 

QC is potentially one of the groundbreaking technologies that can drastically change the picture, in both experimental and theoretical HEP. Possible solutions have been sketched in a series of recent papers, which show the power of QC in areas including computationally heavy algorithms~\cite{Bapst2020,Tuysuz2020}, improved selection of physics signatures~\cite{Mott2017, Felser2021, Wu2021} and more efficient and precise theoretical modelling~\cite{Bauer2021}. In general, QC seems the natural solution to substitute heavily combinatorial algorithms, like those used in track finding in silicon devices and in the clustering of deeply segmented calorimeter signals. On top of this, the availability of large dimension minimization systems, as those from adiabatic annealing machines, can serve as a drop-in replacement for multiple reconstruction and analysis algorithms, based on $\chi^2$ or likelihood minimization. 
The same tool would be the holy grail if applicable to machine learning training~\cite{Guan2021}. QC algorithms could in principle be usable in online systems, for example in the discrimination between useful and background events (the so-called ``trigger''). On top of the problems  described above, their implementations have presently to cope with the bottleneck of the time-consuming preparation of the initial state and hence seem futuristic~\cite{Mott2017}.

\section{Quantum communication}
\label{sect:quantum_communication}
\begin{figure*}[t]
\centering
\includegraphics[width=0.89 \linewidth]{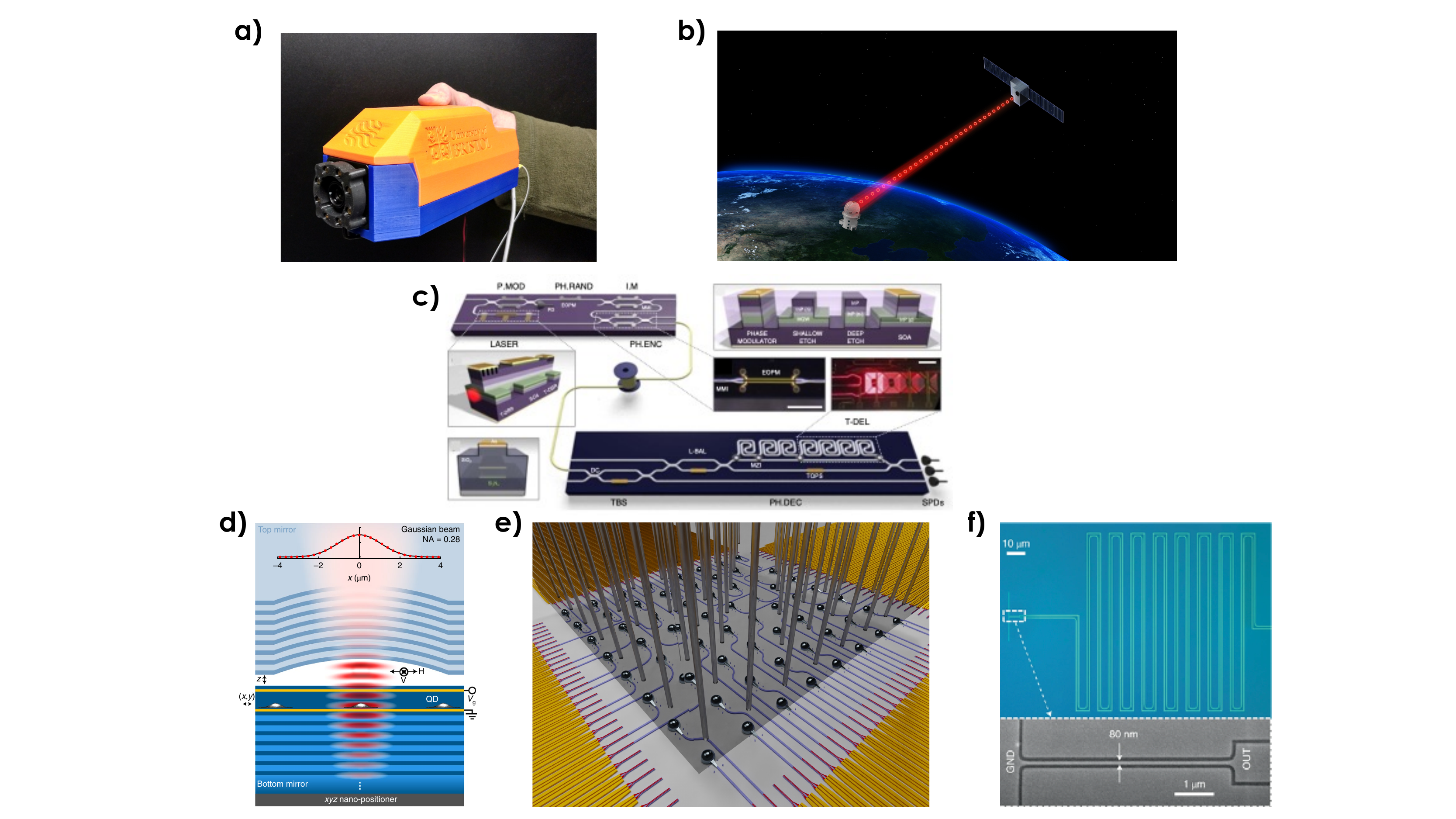}
\caption{{\bf Quantum communications}. The past decade has seen huge interest in ambitious quantum key distribution (QKD) implementations and has spurred the development of quantum devices and platforms. Panel a) Handheld QKD transmitter (courtesy of Dominic O'Brien, joint work of University of Bristol \& University of Oxford, UK quantum communications hub, United Kingdom). Panel b) Satellite QKD linking to long distance fiber optic networks~\cite{Liao2017,Chen2021} (courtesy of Jian-Wei Pan, University of Science and Technology of China). Adapted by permission from Ref.~\cite{Liao2017}. Copyright~\textcopyright~Springer~Nature~(2017). Panel c) Chip-to-chip QKD with integrated photonics~\cite{Sibson2017}. Adapted from Ref.~\cite{Sibson2017} under the terms of the CC 4.0 license (\href{http://creativecommons.org/licenses/by/4.0}{http://creativecommons.org/licenses/by/4.0}). Panel d) High efficiency semiconductor quantum dot coherent single-photon emitter~\cite{Tomm2021}. Adapted by permission from Ref.~\cite{Tomm2021}. Copyright~\textcopyright~Springer~Nature~(2021). Panel e) QKD receiver with waveguide integrated superconducting nanowire single-photon detectors (SNSPDs) (courtesy of Robin Terhaar, University of M\"{u}nster, Germany). Panel f) Ultra low jitter SNSPD~\cite{Korzh2020}. Adapted by permission from Ref.~\cite{Korzh2020}. Copyright~\textcopyright~Springer~Nature~(2020).\label{fig:two}}
\end{figure*}

Quantum communication (Fig.~\ref{fig:two}) is arguably one of the most mature sectors of quantum technology. Progress has been advanced by extensive collaborations across academia and industry. Quantum communication deals with the possibility of exploiting effects such as entanglement, squeezing, and, more generally, coherent superpositions of quantum states, to improve signaling processes between distant parties both in terms of transmission efficiency and security~\cite{Bennett1998}. The overarching goal of quantum communication is the realization of an augmented version of the current global communication network (the quantum secure communication network) that would enable one to incorporate some of the most exotic, yet effective protocols ever conceived in quantum information theory. These include the quantum teleportation algorithm~\cite{Bennett1993} which, by exploiting pre-shared entanglement between the sender and the receiver enables them to efficiently exchange the quantum state of a quantum memory, effectively overcoming the limit imposed by the no-cloning theorem~\cite{Wootters1982}; the superdense coding algorithm~\cite{Bennett1992} which, again by exploiting pre-shared entanglement, allows the communicating parties to boost the rate of  information transmission above the threshold one would assign to a transmission line by purely classical considerations; finally, the quantum key distribution (QKD) algorithm, which instead enables distant users to distill a common private key to be employed for the encryption of messages~\cite{Bennett2014}, by carefully exploiting coherent super-positions of input signals that travel over a public line (possibly monitored by eavesdroppers).

Quantum communication has been a major driver for research into novel quantum devices. Considering the canonical QKD link between Alice and Bob over fiber or free space~\cite{Gisin2002}, there are several main components. The source (Alice), the link, and the detector (Bob). The source may comprise a solid state laser attenuated to the single photon level, photon pair sources~\cite{Lu2019a} or single photon emitters (such as semiconductor quantum dots~\cite{Sellenart2017, Tomm2021} or diamond NV centres~\cite{Fedyanin2021} or defects in 2D materials~\cite{Grosso2017, Schuler2020}). 2D materials are very appealing for on-chip quantum devices for quantum integrated photonic circuits, because of the ease and low-cost of integration with the Si and SiN photonic platforms, and because emitters, modulators, and detectors are all based on the same material platform~\cite{Romagnoli2018,Turunen2022}.

The link can comprise low loss optical fibers or free space (in the atmosphere, space, or underwater).  QKD schemes commonly rely on single photon detectors~\cite{Hadfield2009}.  For free space QKD, including recent space-to-ground QKD demonstrations~\cite{Vallone2015,Liao2017}, photon counters such as Si single-photon avalanche diodes (SPADs) are the common choice in the $600$-$900~{\rm nm}$ wavelength range. The optimal transmission window for optical fibers is $1550~{\rm nm}$ (losses as low as $0.15~{\rm dB}$ per ${\rm km}$ in silica fibers) spurring development of InGaAs SPADs or low noise, low jitter superconducting nanowire single-photon detectors (SNSPDs)~\cite{Korzh2020,Zadeh2021}. Recently transmission at longer (mid-infrared) wavelengths has been considered, for satellite or photonic crystal fiber QKD~\cite{Prabhakar2020}.  As QKD architectures become more mature and ambitious, there are requirements for materials development in other related areas: integrated photonics for on-chip QKD~\cite{Sibson2017,Paraiso2021,Elshaari2020,Beutel2021,Pelucchi2021}, hybrid on-chip integration, random number generators based on quantum devices, high-performance low-loss interconnects, modulators, filters and switches, quantum memories and repeaters.
A major imperative for wider rollout of quantum communications is the threat to conventional public key encryption methods (such as the RSA algorithm~\cite{Rivest1978}) posed by the near-term realization of practical quantum computers capable of implementing powerful factorization algorithms~\cite{Shor1999}. Sensitive governmental, financial or personal data from past or future communications is under threat of capture and decryption. We anticipate there will be increasing demand for mass produced secure quantum communications platforms for scenarios such as satellite-to-ground communications, handheld devices, fiber optic networks and future data centres. Therefore, there is little doubt quantum communication is likely to remain a catalyst for quantum device research for the coming decades.

\section{Quantum sensing platforms and interdisciplinary applications}
\label{sect:quantum_sensing}
\begin{figure*}[t]
\centering
\includegraphics[width=0.89 \linewidth]{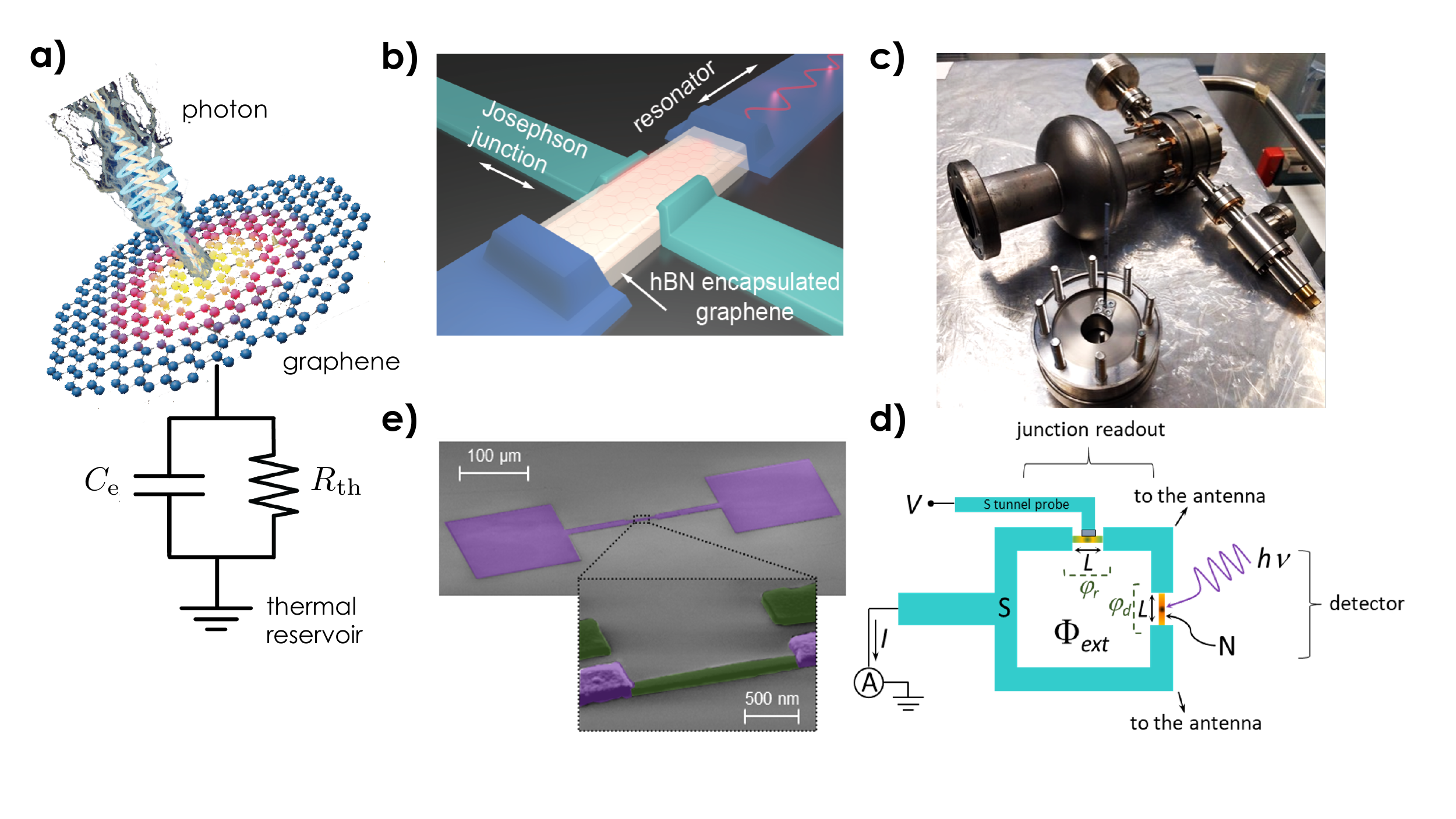}
\caption{{\bf Novel quantum sensing platforms}. 
Panel a) Schematic diagram of a bolometer/calorimeter, working as a single-photon detector. Input photons can heat up the electrons after absorbing into graphene. The thermal energy will dissipate to the thermal reservoir (ground) through the thermal resistance ($R_{\rm th}$), which can be only limited to the electron-phonon coupling channel. When operated as a bolometer, the incident radiation can be inferred by the temperature rise. If the heat capacity ($C_{\rm e}$) is small enough, it can be operated as a calorimeter to detect individual photons as heat pulses with relaxation time on the order of $\sim R_{\rm th}C_{\rm e}$ (courtesy of Graham Rowlands and Minh-Hai Nguyen, Raytheon BBN Technologies, USA). Panel b) Device concept of a graphene-based Josephson junction (GJJ) bolometer~\cite{Lee2020} that is designed to overcome the challenges of both the input coupling of radiation and high-sensitivity detector readout, when applying a new material, i.e.~graphene, for quantum sensing. The Josephson junction, comprising superconductor-graphene-superconductor hybrid system and functioning as a high-sensitivity readout of the graphene electron temperature, is embedded in the current antinode of a microwave resonator to maximize the input coupling. Adapted by permission from Ref.~\cite{Lee2020}. Copyright~\textcopyright~Springer~Nature~(2020).
Panel c) Ultra-high quality factor niobium resonator setup as sample host to study coherence of dielectrics in the quantum regime. Panel d) Temperature-to-phase converter detector: Scheme of a dc SQUID realized with two superconductor(S)/normal metal(N)/superconductor(S) Josephson junctions~\cite{Virtanen2018}. The electronic temperature $T_{\rm e}$ in the right junction is elevated by the absorption of electromagnetic radiation at energy $h\nu$, leading to a suppression of the supercurrent circulating in the interferometer loop. This suppression yields a change of the phase drop ($\varphi_{{\rm d}, {\rm r}}$) across the Josephson junctions, and to a modification of the density of states (DOSs) in the N regions. The DOSs in the readout junction is probed via a superconducting tunnel junction. The device therefore operates as a temperature-to-phase transducer, converting $\delta T_{\rm e} \to \delta\varphi_{{\rm  d}, {\rm r}}$. $V$ is the biasing voltage whereas $I$ is the current flowing through the readout junction. $L$ represents the length of the junctions, and $\Phi_{\rm ext}$ the external magnetic flux piercing the superconducting loop. Adapted with permission from Ref.~\cite{Virtanen2018}. Copyright~\textcopyright~American~Physical~Society~(2018). Panel e) Prototype of a Josephson escape sensor (JES)~\cite{Paolucci2020}. Pseudo-color scanning electron micrograph of a JES. It consists of a 1D fully superconducting nanowire made of an Al/Cu bilayer (green region, as shown in the blow up), connected to two large Al electrodes (violet regions). The proximity Al/Cu bilayer is exploited to precisely tune the intrinsic critical temperature of the nanowire. The typical wire length is $\sim~1~{\rm \mu m}$. Owing to the absorption of radiation, the temperature in the JES is elevate, thereby triggering the transition from the dissipationless to the resistive state. This is similar to what occurs in a conventional transition-edge sensor (TES) although, in contrast to a TES, the JES benefits from an {\it in~situ} fine tuning of its working temperature and sensitivity thanks to a current flowing on purpose through the nanowire.\label{fig:three}}
\end{figure*}

Exploiting their sensitivity to the external environment, quantum sensors can outperform detectors that rely on classical principles alone~\cite{Degen2017} (see Fig.~\ref{fig:three}). Examples such as superconducting interference devices, atomic systems, and NV centers in diamond~\cite{Bucher2019,Casola2018,Yang2020,Hart2021} have demonstrated how quantum phenomena can achieve unprecedented sensitivities in measuring time, electric, and magnetic fields. 

More recently, quantum sensing is becoming more interdisciplinary with applications going well beyond quantum information science and traditional areas of condensed matter physics. An exciting frontier is that of applications to astroparticle physics, i.e.~to the search of dark matter axions via magnetic topological insulators~\cite{Marsh2019}, Josephson parametric amplifiers~\cite{Bartram2021,Lehnert2021}, and superconducting qubits~\cite{Dixit2021}. 

Materials research will continue to advance quantum sensing by improving coherence and suppressing losses, as well as to expand applications to a broader range.

\subsection{Superconducting nanowire single-photon detectors, transition edge sensors, and microwave kinetic inductance detectors}

In the near-infrared regime, superconducting single-photon detectors (SPD), such as SNSPDs (see Sect.~\ref{sect:quantum_communication}), transition edge sensors (TESs), and microwave kinetic inductance detectors (MKIDs) have been key building blocks for quantum technologies, i.e.~for revolutionary free-space quantum communication protocols~\cite{Chen2021}. Popular choices of materials for these detectors include~\cite{Mazin2021} Al, Nb, MoSi, WSi, NbN, and NbTiN. Further studies on the details of the underlying detection mechanisms and superconductor properties~\cite{Frasca2019, Zhao2020b} are shedding light on how to optimize the performance of these detectors. For example, they will teach us how to operate at higher temperatures, i.e.~by capitalizing on the recent efforts employing materials like ${\rm MgB}_2$, bismuth strontium calcium copper oxide (BSCCO), and YBCO~\cite{Shibata2010, Lyatti2020}, or at lower photon energies, i.e.~in the mid- and far-infrared spectral regions, 
where the number of quasiparticles generated by a single photon decreases.

\subsection{vdW sensors} 

In the new frontier of single-photon detection in the THz and microwave regimes, vdW materials are emerging as a promising platform for quantum sensing~\cite{Liu2019}. These materials can be only one or a few atomic layers thick, with ultrahigh photoconductive gain~\cite{Konstantatos2012} and minute thermal conductance, resulting in a relatively large bolometric temperature rise upon photon absorption. Specifically, for graphene, its electron heat capacity can be vanishingly small near the charge neutrality point due to its pseudo-relativistic band structure. By overcoming the challenge of photon coupling and thermal readout based on the Josephson effect~\cite{Walsh2021}, it has been demonstrated that the sensitivity of graphene bolometers can reach the fundamental limits imposed by thermal fluctuations~\cite{Lee2020},  suitable for quantum circuits~\cite{Kokkoniemi2020}. Single-photon detection of near-infrared photons has recently been demonstrated using a graphene-based Josephson junction~\cite{Walsh2021}. Due to graphene’s capability of coupling to electromagnetic waves in a wide range of frequencies, the fast thermalization time of its carriers, and its ultrafast response, graphene-based detectors represent an attractive solution for broadband quantum sensing, provided that device fabrication will be improved and made scalable.

Finally, as we will discuss in Sect.~\ref{sect:quantum_simulation}, we highlight that moir\'{e} superconductors created in vdW heterostructures offer interesting opportunities for quantum sensing. Indeed, these systems have large kinetic inductances that can be useful for the development of MKIDs or TESs~\cite{Seifert2020}, since the superconducting carrier density can be lower than conventional superconductors by $5$ orders of magnitude.

\subsection{Niobium resonators} 
The recent advances---discussed above in Sect.~\ref{sect:superconducting cavities}---in 3D cQED architectures and in improving the coherence of SRF cavities in the quantum regime~\cite{Romanenko2020} pave the way for important applications also in the areas of quantum sensing and metrology. Recent experiments have indeed demonstrated how these very-high-quality-factor resonators offer e.g.~microwave characterization environments of uniquely high sensitivity, allowing one to study, with great precision, the loss tangents of various materials for quantum devices, including dielectrics such as silicon and sapphire~\cite{Checchin2021}. In the area of quantum sensing, important applications include the search for dark sector particles such as dark matter and dark photons~\cite{Peccei1977,Dixit2021}. 

Cavity systems have also enabled the isolation and detailed study of a single electron, providing the most precise measurement of a fundamental property of matter~\cite{Hanneke2008}, now a textbook demonstration of the power of quantum field theory at the part per trillion level. The coherence of the quantum devices employed in this class of experiments is a critical factor in determining the sensitivity of the particle search or property measurement. Important materials advancements and research directions include coating these 3D structures with thin films of superconductors, such as ${\rm Nb}_3{\rm Sn}$, ${\rm NbTi}$ or higher-$T_{\rm c}$ materials such as rare-earth barium copper oxide (also known as ReBCO), that minimize microwave losses in the presence of high magnetic fields~\cite{Golm2021,Alesini2021}, since the presence of a large magnetic field (up to several Teslas) will increase the discovery potential of the experiments.

\subsection{New device concepts and designs}

Some novel superconducting platforms for sensing have been envisioned and experimentally realized in the last few years. Among them, one original approach is to exploit the principle of temperature-to-phase~\cite{Virtanen2018} conversion, which is based on the exploitation of superconductor/normal metal/superconductor DC proximity SQUIDs. Here, absorption of radiation yields heating of one of the two junctions of the SQUID, thereby inducing a change of the phase drop across the other junction, which is detected via a tunnel probe connected to it---see Fig.~\ref{fig:three}d). 

Other alternative device concepts rely on the use of ferromagnetic insulators-based superconducting tunnel junctions exploiting, for instance, thin EuS layers coupled to the superconductor so as to induce a sizable thermoelectric effect in the structure~\cite{Heikkila2018}. Heating one part of the junction due to radiation absorption translates into a thermovoltage output signal, which can then be detected. Still in the context of superconducting tunnel junctions, a single-photon thermal detector operating in the dissipationless regime, and based on the sharp jump of the critical current of a temperature-biased tunnel junction has been proposed~\cite{Guarcello2019}. 

Recently, an Al-based quantum capacitance detector~\cite{Echternach2018}, where photon-excited quasiparticles tunnel into a Coulombic island embedded in a resonant circuit, displayed an ultra-low noise-equivalent power (NEP) below $10^{-20}~{\rm W}/{\rm Hz}^{1/2}$ at $1.5~{\rm THz}$. Finally, a prototype for a Josephson escape sensor (JES) consisting of a fully superconducting 1D Al-based Josephson junction was also demonstrated~\cite{Paolucci2020}---see Fig.~\ref{fig:three}e). The JES is expected to yield thermal fluctuation noise-limited NEP around $10^{-25}~{\rm W}/{\rm Hz}^{1/2}$ if operated as a bolometer, and a frequency resolution of about $2~{\rm GHz}$ when operated as a calorimeter.

\section{Solid-state quantum simulators based on moir\'{e} quantum matter}
\label{sect:quantum_simulation}

Interactions between matter constituents, such as electrons in a crystal, are the source of some of the most fascinating states of matter, such as strange metals, exotic superconductors, and topologically-ordered phases with fractionalized excitations. Programmable quantum simulators~\cite{Altman2021} are ideal candidates for the accurate investigation of these strongly correlated quantum phases. For example, the onset of a quantum spin liquid phase of the paradigmatic toric code~\cite{Kitaev2003} type was detected in 2D arrays of $219$ $^{87}{\rm Rb}$ atoms individually trapped in optical tweezers and positioned on the links of a kagome lattice~\cite{Semeghini2021}. Similarly, toric code states have been prepared and analyzed by using a quantum circuit comprising $31$ superconducting qubits~\cite{Satzinger2021}. 
While these systems are highly controllable and can faithfully realize certain quantum Hamiltonians, engineering tunable long-range interactions remains an outstanding experimental challenge. For example, two atoms in Rydberg states interact with each other through a vdW potential decaying like the inverse sixth power of the relative distance. On the contrary, the relevant e-e interaction potential in solids decays like the inverse first power of the relative distance (Coulomb law). The quantum simulation of a plethora of strongly correlated phases of matter therefore requires controllable and ultra-clean quantum many-particle systems interacting through Coulomb forces. 

\begin{figure*}[t]
\centering
\includegraphics[width=0.89 \linewidth]{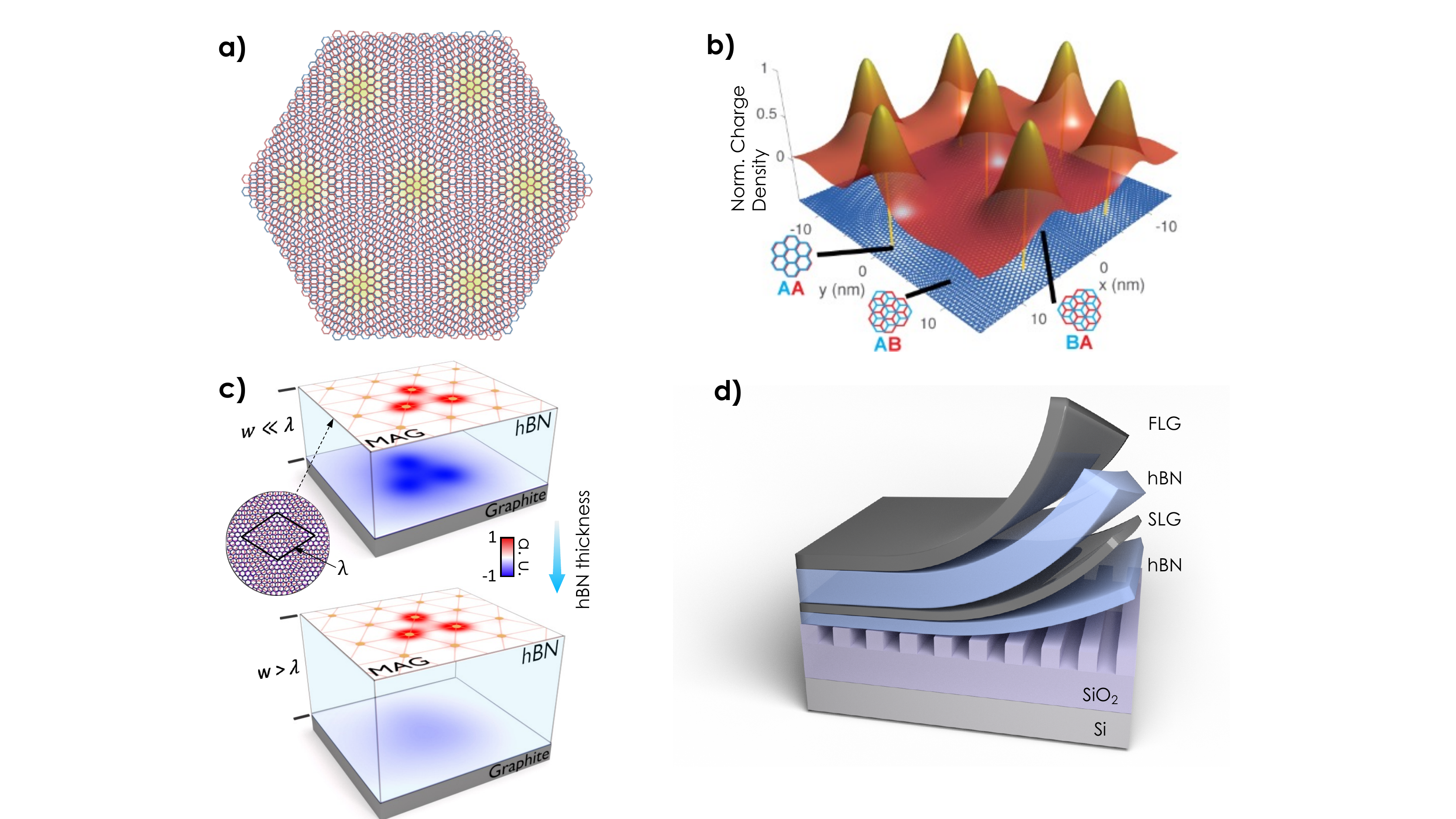}
\caption{{\bf Natural and synthetic moir\'e quantum matter for quantum simulation}. Panel a) A moir\'{e} pattern created by a relative rotation of two 2D honeycomb lattices with identical lattice spacing. Adapted by permission from Ref.~\cite{Cao2018a}. Copyright~\textcopyright~Springer~Nature~(2018). Panel b) Normalized local density of states calculated for twisted bilayer graphene (TBG) flat bands with positive energy at a twist angle $\theta=1.08\degree$. The electron density is strongly concentrated at the regions with AA stacking order, whereas it is mostly depleted at AB- and BA-stacked regions. Adapted by permission from Ref.~\cite{Cao2018a}. Copyright~\textcopyright~Springer~Nature~(2018). Panel c) Correlations in TBG can be altered by a series of tuning parameters. One of those is screening from a nearby ultra-flat metallic gate such as a graphite gate placed at a distance $w$ from e.g.~magic-angle twisted bilayer graphene (MAG). By increasing the thickness of the hexagonal Boron Nitride (hBN) spacer, one can investigate samples where $w\ll \lambda$ or $w> \lambda$, $\lambda$ being the wavelength of the moir\'e superlattice. Adapted by permission from Ref.~\cite{Stepanov2020}. Copyright~\textcopyright~Springer~Nature~(2020). Panel d) Cartoon schematic showing the architecture of a 1D patterned dielectric superlattice graphene device structure. The superlattice potential arises due to patterning of the ${\rm SiO_2}$ dielectric layer that separates the highly doped Silicon substrate (functioning as a metallic back gate) from the hBN-encapsulated single-layer graphene (SLG) device. Few-layer graphene (FLG) can be used as a top metallic gate. Of course, much more complicated patterns can be created, with the aim of realizing synthetic MQM with flat bands~\cite{Li2021} (courtesy of Cory Dean, Columbia University, USA). Adapted by permission from Ref.~\cite{Li2021}. Copyright~\textcopyright~Springer~Nature~(2021).\label{fig:four}}
\end{figure*}

In this respect, a breakthrough is represented by the discovery of correlated insulators and superconductivity in magic-angle twisted bilayer graphene~\cite{Cao2018a, Cao2018b}. These findings have led to the emergence of a new platform, which has been dubbed moir\'{e} quantum matter~\cite{Andrei2021} (MQM), to investigate strongly correlated and topological physics of 2D electron systems (see Fig.~\ref{fig:four}). MQM refers to long-wavelength superlattices obtained by stacking or rotating relatively to each other two or more 2D crystals such as graphene, hexagonal boron nitride, and transition metal dichalcogenides. A periodic modulation of the inter-layer electron hopping in these stacks leads to band flattening~\cite{Bistritzer2011} (bandwidth of the flat bands on the order of $10$-$20~{\rm meV}$), which, in turn, results into strong correlations (on-site Hubbard interaction energy scale on the order of $50$-$100~{\rm meV}$). 

The MQM platform has led to unprecedented electrical control over multiple phases of matter in 2D material heterostructures. Unraveling the exact microscopic mechanisms that are responsible for superconducting phases in MQM is currently one of the biggest challenges in the field. In these systems, superconductivity is believed to be unconventional, as it often arises in the presence of strong electronic correlations~\cite{Cao2018b} or close to magnetically ordered states~\cite{Lu2019}. In magic-angle twisted trilayer graphene~\cite{Park2021, Zao2021} the superconducting phase has been explicitly shown to violate the Pauli limit~\cite{Cao2021}, which likely points towards a spin-triplet Cooper pairing. The experimental exploration of collective modes in these materials is still at its infancy~\cite{Hesp2021}. As sample quality increases and tools like cryogenic near-field optical spectroscopy~\cite{Ni2018, Sun2020} become more and more developed, several new types of collective excitations are expected to be discovered and linked with broken symmetry phases of MQM.

Since MQM is made from stacks of 2D materials, it possesses revolutionary tuning parameters that can be used a) to accurately control interaction energy scales and symmetries for the simulation of fundamental physics problems~\cite{Kennes2021} and b) to enable entirely new device concepts for quantum technologies. As far as point a) is concerned, for example, inter-layer alignment between the various layers can be used to alter symmetries and band widths~\cite{Sharpe2019}. Metallic electrodes (i.e.~gates) can be used to control the electron filling via the electric field effect and alter the range of e-e interactions via screening~\cite{Stepanov2020}. With respect to point b), we mention for example ``hyper-tunable'' Josephson junctions~\cite{RodanLegrain2021, deVries2021} (where both the superconductor and weak link are realized within the same moir\'{e} material and they are both electrically tunable) and exquisitely sensitive quantum photodetectors~\cite{Seifert2020, Deng2020, DiBattista2021}, with unprecedented capabilities to detect single photons in the Terahertz and microwave regimes. 
A series of important steps have to be made by the community, though, to make sure that MQM becomes an established platform for quantum simulation. First, twist-angle inhomogeneities~\cite{Uri2020} have to be tamed. Second, new methods to assemble and nanofabricate MQM on large scales are needed. While some robotic and machine-learning enhanced automated fabrication methods for 2D materials have been developed recently~ \cite{Masubuchi2018, Han2020}, substantial efforts are needed to extend them to precisely controlled moir\'{e} heterostructures and to achieve large-scale fabrication capabilities.

An alternative approach to realize MQM consists in placing single- or few-layer graphene over a dielectric~\cite{Drienovsky2018} or metallic~\cite{Song2018} substrate which has been patterned. Such periodic spatial modulations can lead to a synthetic version of MQM that is in principle more versatile than the ``natural'' MQM discussed above, with freedom to choose crystal symmetry, synthetic strain, etc. Replica Dirac cones, Hofstadter butterfly spectra, and anisotropic electron transport have already been demonstrated via patterning techniques~\cite{Forsythe2018, Huber2020, Li2021}. The holy grail of synthetic MQM is to realize highly tunable flat electronic bands~\cite{Shi2019} such that strong e-e interactions can be induced. This enables the exploration of correlated phenomena such as superconductivity and magnetism with a highly controllable system.

\section{Quantum technologies for the energy sector}
\label{sect:energy}
A great deal of interest is currently also nucleating around quantum technologies for the energy sector~\cite{Ho2018,Ajagekar2019}. Many researchers in the past decade have studied the impact of quantum effects on a variety of devices and systems (including biological and bio-artificial ones) for the harvest and transmission of energy. For example, quantum coherence and interference effects have been proposed as means to suppress charge recombination losses and enhance the efficiency of photovoltaic devices~\cite{Scully2011,Bittner2014} as well as to increase the yield of photosynthetic architectures~\cite{Chin2013,Dorfman2013}. Similarly, several authors have discussed the possibility to boost the performance of organic photovoltaic cells by relying on long-range energy transfer between organic molecules enabled by vacuum-field cavity polaritons~\cite{Zhong2016,Zhong2017,Rozenman2018,SaezBlazquez2018}.
Nowadays, however, the first thing that comes to mind, when the words ``quantum'' and ``energy'' are mentioned together, is probably applications of QC to computational quantum chemistry and materials science (see Sect.~\ref{sect:quantum_computing_applications}). Obvious and timely directions here are those related to the increase of the performance of electrochemical batteries. An example is that of scientists from IBM and Mitsubishi Chemical, who have simulated on a NISQ device the initial steps of the reaction mechanism between lithium and oxygen in lithium-air batteries~\cite{Rice2021,Gao2021}.

But clearly there is much more than that. How can quantum mechanics help breakthroughs in energy technology? We foresee three possible groundbreaking directions. 

a) First, solid-state devices whose functionalities are enabled by quantum mechanical resources such as entanglement and coherence~\cite{Fornieri2017, Benenti2017, Pekola2021} will play a much broader role in the future. Indeed, it is difficult to imagine a ``quantum society'' where computing, communications, and sensing will be carried out by relying on the principles of quantum mechanical laws while energy will be provided by systems ruled by standard electrochemical laws and 200-years old designs. 
There are three classes of energy-related quantum devices, which are currently being mainly studied by academia and that will need to gradually be evaluated and adopted by industry: i) quantum heat engines for energy production~\cite{Josefsson2018,vonLindenfels2019,Klatzow2019,Prete2019,Ono2020, Scharf2020, Marchegiani2020, Guthrie2021}, ii) quantum batteries~\cite{Alicki2013,Hovhannisyan2013,Binder2015, Campaioli2017,Campaioli2018,Ferraro2018,Andolina2019,Sergi2020,Rossini2020,Quach2020,Hu2021} and supercapacitors~\cite{Ferraro2019} for energy storage, and iii) quantum energy lines~\cite{Tirone2021} for energy transfer. Research and development on all these aspects may lead to a profound ``cultural'' and technological revolution in how energy is produced, stored, and conveyed.

b) Second, we believe that also cavity quantum electrodynamical control of the ground and excited states of molecular systems and quantum phase transitions of extended electron systems~\cite{GarciaVidal2021,Hubener2021} may play an important role in the energy sector. When electromagnetic fields are confined to a cavity, vacuum fluctuations are greatly enhanced and are expected to alter the properties of a variety of quantum systems. Experimentally, it has been shown for example that the vacuum field can favor a particular product in transformations of compounds that have two or more reactive sites, provided that vibrational strong coupling between a reactant and the vacuum field of an optical cavity is achieved~\cite{Thomas2019}. Experimental evidence of the impact of vacuum fields on superconducting compounds~\cite{Thomas2021} and 2D electron gas physics~\cite{Paravicini2019} has also been reported. If and how quantum fluctuations of cavity electromagnetic fields will play a pivotal role in energy-related issues is certainly an interesting research direction. For example, strong coupling between electronic degrees of freedom and the vacuum field has been shown to lead to a large increase in the capacitance of a quantum supercapacitor~\cite{Ferraro2019}.  

c) Third, quantum mechanics can also help to develop green quantum technologies. For example, qubit driving schemes that reuse and correct drive pulses can help reduce the average energy consumption per gate operation in a quantum computer, without increasing the average gate error~\cite{Ikonen2017}. Ingenious QC schemes where energy is recycled from one computational step to the next have also been proposed~\cite{Chiribella2021}. Finally, constraints on the physical resources for quantum computation have been shown~\cite{Fellous2021} to lead to a maximum probability that an error occurs in a physical gate or qubit that grows as the scale of the quantum computer grows. Energy minimization constraints are an obviously important resource that leads to such scale-dependent noise. A full energetic optimization of a large-scale quantum computer that includes optimizing cryogenics, control circuitry, signal generation, and the quantum algorithm for the calculation at hand will be needed for the development of greener future quantum technologies.

\begin{acknowledgments}
We wish to thank Matvey Lyatti for preparing the bottom right panel of Fig.~\ref{fig:one}, Dominic O'Brien for sending us Fig.~\ref{fig:two}a), Jian-Wei Pan for sending us Fig.~\ref{fig:two}b), Robin Terhaar for preparing Fig.~\ref{fig:two}e), Graham Rowlands and Minh-Hai Nguyen for preparing Fig.~\ref{fig:three}a), Cory Dean for sending us Fig.~\ref{fig:four}d), and Ettore Vicari for useful discussions. 

M.P. and F.H.L.K. are supported by the European Union's Horizon 2020 research and innovation programme under Grant Agreement No.~881603 -- GrapheneCore3.

F.G. is supported by the European Research Council (ERC) under Grant Agreement No.~899315 -- TERASEC and, partially, by the EU’s  Horizon 2020 research and innovation program under Grant Agreement No.~800923 (SUPERTED) and No.~964398 (SUPERGATE). 

K.C.F. acknowledges support from the Army Research Office under Contract W911NF-18-C-0044. 

C.S. acknowledges the support from the Ministry for Culture and Science of North Rhine-Westphalia (No.~421-8.03.03.02-130428). 

G.S. acknowledges the support from INFN under the projects LITEBIRD and LSPE, and from the Italian Space Agency (ASI) under grants No.~2020-9-HH.0 and 2020-25-HH.0. 

M.D. and M.P. acknowledge support from the University of Pisa under the ``PRA - Progetti di Ricerca di Ateneo" (Institutional Research Grants) -  Project No.~PRA\_2020-2021\_92 ``Quantum Computing, Technologies and Applications''. 

R.H.H. is supported by UK research and innovation (UKRI) through Grants EP/S026428/1, EP/T001011/1, EP/T00097X/1 and ST/T006331/1. 

V.G. is supported by MIUR (Ministero dell'Istruzione, dell'Universit\`{a} e della Ricerca) via project PRIN 2017 ``Taming complexity via QUantum Strategies: A Hybrid Integrated Photonic approach'' (QUSHIP), grant no.~2017SRNBRK. 

D.K.E. acknowledges support from the Ministry of Economy and Competitiveness of Spain through the ``Severo Ochoa'' program for Centres of Excellence in R\&D (SE5-0522), Fundaci\'{o} Privada Cellex, Fundaci\'{o} Privada Mir-Puig, the Generalitat de Catalunya through the CERCA program, funding from the ERC under the European Union’s Horizon 2020 research and innovation programme (grant agreement No.~852927).

F.H.L.K. also acknowledges support from the ERC under Grant Agreement No.~726001 -- TOPONANOP, the government of Spain (PID2019-106875GB-I00; Severo Ochoa CEX2019-000910-S [MCIN/ AEI/10.13039/501100011033]), Fundaci\'{o} Cellex, Fundaci\'{o} Mir-Puig, and Generalitat de Catalunya (CERCA, AGAUR, SGR 1656).

P.J.H. acknowledges support by the National Science Foundation QII-TAQS grant no.~OMA-1936263, the U.S. Department of Energy, Office of Science, Basic Energy Sciences, under Award Number DE-SC0020149, the Center for the Advancement of Topological Semimetals, an Energy Frontier Research Center funded by the U.S. Department of Energy Office of Science, through the Ames Laboratory under contract DE-AC02-07CH11358, the Army Research Office through grant No.~W911NF1810316, the US Department of Energy (DOE), Office of Basic Energy Sciences (BES), Division of Materials Sciences and Engineering under Award DESC0001819, the National Science Foundation (DMR-1809802), the STC Center for Integrated Quantum Materials (NSF Grant No. DMR-1231319), the Air Force Office of Scientific Research (AFOSR) 2DMAGIC MURI FA9550-19-1-0390, the Gordon and Betty Moore Foundation’s EPiQS Initiative through Grant GBMF9643, the Fundación Ramon Areces and the CIFAR Quantum Materials program. 

The work of A.G. is supported by the U.S. Department of Energy, Office of Science, National Quantum Information Science Research Centers, Superconducting Quantum Materials and Systems Center (SQMS) under contract number DE-AC02-07CH11359. 
\end{acknowledgments}


\begin{thebibliography}{77}
%
\bibitem{Aspect2003}
 Aspect, A. John Bell and the second quantum revolution. In Bell, J. S. \href{https://doi.org/10.1017/CBO9780511815676}{{\it Speakable and unspeakable in quantum mechanics: Collected papers on quantum philosophy}} (Cambridge University Press, Cambridge, 2004). 
%
\bibitem{Dowling2003}
Dowling, J. P. \& Milburn, G. J. 
Quantum technology: The second quantum revolution.
\href{https://doi.org/10.1098/rsta.2003.1227}{{\it Phil. Trans. R. Soc. Lond. A}~{\bf 361}, 1655 (2003)}.
%
\bibitem{Deutsch2020}
Deutsch, I. H. Harnessing the power of the second quantum revolution. \href{https://doi.org/10.1103/PRXQuantum.1.020101}{{\it PRX Quantum}~{\bf 1}, 020101 (2020)}.
%
\bibitem{Raymer2019}
Raymer, M. G. \& Monroe, C. The US national quantum initiative. \href{https://doi.org/10.1088/2058-9565/ab0441}{{\it Quantum Sci. Technol.}~{\bf 4}, 020504 (2019)}
%
\bibitem{Knight2019}
Knight, P. \& Walmsley, I. UK national quantum technology programme. \href{https://doi.org/10.1088/2058-9565/ab4346}{{\it Quantum Sci. Technol.}~{\bf 4}, 040502 (2019)}.
%
\bibitem{Riedel2019}
Riedel, M., Kovacs, M., Zoller, P., Mlynek, J. \& Calarco, T. Europe's quantum flagship initiative. \href{https://doi.org/10.1088/2058-9565/ab042d}{{\it Quantum Sci. Technol.}~{\bf 4}, 020501 (2019)}.
%
\bibitem{Zhang2019_Intro}
Zhang, Q., Xu, F., Li, L., Liu, N.-L. \& Pan, J.-W.
Quantum information research in China. \href{https://doi.org/10.1088/2058-9565/ab4bea}{{\it Quantum Sci. Technol.}~{\bf 4}, 040503 (2019)}.
%
\bibitem{Schafer2020} 
Sch\"{a}fer, F., Fukuhara, T., Sugawa, S., Takasu, Y. \&  Takahashi, Y. Tools for quantum simulation with ultracold atoms in optical lattices. \href{https://doi.org/10.1038/s42254-020-0195-3}{{\it Nat. Rev. Phys.}~{\bf 2}, 411 (2020)}.
%
\bibitem{Monroe2021} 
Monroe, C., Campbell, W. C., Duan, L.-M., Gong, Z.-X., Gorshkov, A. V.,  Hess, P., Islam, R., Kim, K., Pagano, G., Richerme, P., Senko, C. \& Yao, N. Y. Programmable quantum simulations of spin systems with trapped ions. 
\href{https://doi.org/10.1103/RevModPhys.93.025001}{{\it Rev. Mod. Phys.}~{\bf 93}, 025001 (2021)}.
%
\bibitem{Adams2020} 
Adams, C. S., Pritchard, J. D. \& Shaffer, J. P. Rydberg atom quantum technologies. \href{https://doi.org/10.1088/1361-6455/ab52ef}{{\it J. Phys. B: At. Mol. Opt. Phys.}~{\bf 53}, 012002 (2020)}.
%
\bibitem{Paik2011}
Paik, H., Schuster, D. I.,  Bishop, L. S., Kirchmair, G., Catelani, G., Sears, A. P., Johnson, B. R.,  Reagor, M. J., Frunzio, L., Glazman, L. I., Girvin, S. M., Devoret, M. H. \& Schoelkopf, R. J. Observation of high coherence in Josephson junction qubits measured in a three-dimensional circuit QED architecture. \href{https://doi.org/10.1103/PhysRevLett.107.240501}{{\it Phys. Rev. Lett.}~{\bf 107}, 240501 (2011)}.
%
\bibitem{Reagor2016}
Reagor, M., Pfaff, W.,  Axline, C., Heeres, R. W., Ofek, N., Sliwa, K., Holland, E.,  Wang, C.,  Blumoff, J.,  Chou, K., Hatridge, M. J., Frunzio, L., Devoret, M. H., Jiang, L. \& Schoelkopf, R. J. Quantum memory with millisecond coherence in circuit QED. \href{https://doi.org/10.1103/PhysRevB.94.014506}{{\it Phys. Rev. B}~{\bf 94}, 014506 (2016)}.
%
\bibitem{Blais2021}
Blais, A., Grimsmo, A. L., Girvin, S. M. \& Wallraff, A. Circuit quantum electrodynamics. \href{https://doi.org/10.1103/RevModPhys.93.025005}{{\it Rev. Mod. Phys.}~{\bf 93}, 025005 (2021)}.
%
\bibitem{Romanenko2020} 
Romanenko, A, Pilipenko, R., Zorzetti, S., Frolov, D., Awida, M, Belomestnykh, S., Posen, S. \& Grassellino, A. Three-dimensional superconducting resonators at $T<20~{\rm mK}$ with photon lifetimes up to $\tau = 2~{\rm s}$. \href{https://doi.org/10.1103/PhysRevApplied.13.034032}{{\it Phys. Rev. Appl.}~{\bf 13}, 034032 (2020)}.
%
\bibitem{Vlastakis2013} 
Vlastakis, B., Kirchmair, G., Leghtas, Z., Nigg, S. E., Frunzio, L., Girvin, S. M., Mirrahimi, M., Devoret, M. H. \& Schoelkopf, R. J. Deterministically encoding quantum information using 100-photon Schr\"{o}dinger cat states. \href{https://doi.org/10.1126/science.1243289
}{{\it Science}~{\bf 342}, 607 (2013)}. 
%
\bibitem{Ofek2016} 
Ofek, N., Petrenko, A., Heeres, R., Reinhold, P., Leghtas, Z., Vlastakis, B., Liu, Y., Frunzio, L.,  Girvin, S. M., Jiang, L., Mirrahimi, M., Devoret, M. H. \& Schoelkopf, R. J. Extending the lifetime of a quantum bit with error correction in superconducting circuits. \href{https://doi.org/10.1038/nature18949}{{\it Nature}~{\bf 536}, 7617 (2016)}. 
%
\bibitem{Terhal2020} 
Terhal, B. M., Conrad, J. \& Vuillot, C. Towards scalable bosonic quantum error correction. 
\href{https://doi.org/10.1088/2058-9565/ab98a5}{{\it Quantum Sci. Technol.}~{\bf 5}, 043001 (2020)}.
%
\bibitem{Cai2021}
Cai, W., Ma, Y., Wang, W., Zou, C. L. \& Sun, L. Bosonic quantum error correction codes in superconducting quantum circuits. \href{https://doi.org/10.1016/j.fmre.2020.12.006}{{\it Fund. Res.}~{\bf 1}, 50 (2021)}.
%
\bibitem{Naik2017}
Naik, R. K., Leung, N., Chakram, S., Groszkowski, P., Lu, Y., Earnest, N., McKay, D. C., Koch, J. \& Schuster, D. I. Random access quantum information processors using multimode circuit quantum electrodynamics. \href{https://doi.org/10.1038/s41467-017-02046-6}{{\it Nat. Commun.}~{\bf 8}, 1904 (2017)}.
%
\bibitem{Satzinger2021}
Satzinger, K. J., Liu, Y., Smith, A., Knapp, C., Newman, M., Jones, C., Chen, Z., Quintana, C., Mi, X., Dunsworth, A., Gidney, C., Aleiner, I., Arute, F., Arya, K., Atalaya, J., Babbush, R., Bardin, J. C., Barends, R., Basso, J., Bengtsson, A., Bilmes, A., Broughton, M., Buckley, B. B., Buell, D. A., Burkett, B., Bushnell, N., Chiaro, B., Collins, R., Courtney, W., Demura, S., Derk, A. R., Eppens, D., Erickson, C., Farhi, E., Faoro, L., Fowler, A. G., Foxen, B., Giustina, M., Greene, A., Gross, J. A., Harrigan, M. P., Harrington, S. D., Hilton, J., Hong, S., Huang, T., Huggins, W. J., Ioffe, L. B.,  Isakov, S. V., Jeffrey, E., Jiang, Z., Kafri, D., Kechedzhi, K., Khattar, T., Kim, S., Klimov, P. V.,  Korotkov, A. N., Kostritsa, F., Landhuis, D, Laptev, P., Locharla, A., Lucero, E., Martin, O., McClean, J. R., McEwen, M., Miao, K. C., Mohseni, M., Montazeri, S., Mruczkiewicz, W., Mutus, J., Naaman, O., Neeley,  M., Neill, C., Niu, M. Y., O'Brien, T. E., Opremcak, A, Pat\'{o}, B., Petukhov, A., Rubin, N. C., Sank, D., Shvarts, V., Strain, D., Szalay, M., Villalonga, B., White, T. C., Yao, Z., Yeh, P., Yoo, J., Zalcman, A., Neven, H., Boixo, S., Megrant, A., Chen, Y., Kelly, J., Smelyanskiy, V., Kitaev, A., Knap, M., Pollmann, F. \& Roushan, P. Realizing topologically ordered states on a quantum processor. 
\href{https://doi.org/10.1126/science.abi8378}{{\it Science}~{\bf 374}, 1237 (2021)}.
%
\bibitem{Niepce2019} 
Niepce, D., Burnett, J. \& Bylander, J. High kinetic inductance NbN nanowire superinductors. 
\href{https://doi.org/10.1103/PhysRevApplied.11.044014}{{\it Phys. Rev. Appl.}~{\bf 11}, 044014 (2019)}.
%
\bibitem{Hazard2019}
Hazard, T. M., Gyenis, A., Di Paolo, A., Asfaw, A. T., Lyon, S. A., Blais, A. \& Houck, A. A. 
Nanowire superinductance fluxonium qubit. 
\href{https://doi.org/10.1103/PhysRevLett.122.010504}{{\it Phys. Rev. Lett.}~{\bf 122}, 010504 (2019)}.
%
\bibitem{Shearrow2018} 
Shearrow, A., Koolstra, G., Whiteley, S. J., Earnest, N., Barry, P. S., Heremans, F. J., Awschalom, D. D., Shirokoff, E. \& Schuster, D. I. Atomic layer deposition of titanium nitride for quantum circuits. \href{https://doi.org/10.1063/1.5053461}{{\it Appl. Phys. Lett.}~{\bf 113}, 212601 (2018)}.
%
\bibitem{Maleeva2018} 
Maleeva, N., Gr\"{u}nhaupt, L., Klein, T., Levy-Bertrand, F., Dupre, O., Calvo, M., Valenti, F., Winkel, P., Friedrich, F., Wernsdorfer, W., Ustinov, A. V., Rotzinger, H., Monfardini, A., Fistul, M. V. \& Pop, I. M.  Circuit quantum electrodynamics of granular aluminum resonators. \href{https://doi.org/10.1038/s41467-018-06386-9}{{\it Nat. Commun.}~{\bf 9}, 3889 (2018)}.
%
\bibitem{Grunhaupt2019} 
Gr\"{u}nhaupt, L., Spiecker, M., Gusenkova, D., Maleeva, N., Skacel, S. T., Takmakov, I., Valenti, F., Winkel, P., Rotzinger, H., Ustinov, A. V. \& Pop, I. M. Granular aluminium as a superconducting material for high-impedance quantum circuits. \href{https://doi.org/10.1038/s41563-019-0350-3}{{\it Nat. Mater.}~{\bf 18}, 816 (2019)}.
%
\bibitem{Winkel2020} 
Winkel, P., Borisov, K., Gr\"{u}nhaupt, L., Rieger, D., Spiecker, M., Valenti, F., Ustinov, A. V.,  Wernsdorfer, W. \& Pop, I. M. Implementation of a transmon qubit using superconducting granular aluminum. \href{https://doi.org/10.1103/PhysRevX.10.031032}{{\it Phys. Rev. X}~{\bf 10}, 031032 (2020)}.
%
\bibitem{Pita-Vidal2020}
Pita-Vidal, M., Bargerbos, A., Yang, C.-K., van Woerkom, D. J., Pfaff, W., Haider, N., Krogstrup, P., Kouwenhoven, L. P., de Lange, G. \& Kou, A. A gate-tunable, field-compatible fluxonium. 
\href{https://doi.org/10.1103/PhysRevApplied.14.064038}{{\it Phys. Rev. Applied}~{\bf 14}, 064038 (2020)}.
%
\bibitem{YangF2020} 
Yang, F., Gozlinski, T., Storbeck, T., Gr\"{u}nhaupt, L., Pop, I. M. \& Wulfhekel, W. Microscopic charging and in-gap states in superconducting granular aluminum. \href{https://doi.org/10.1103/PhysRevB.102.104502}{{\it Phys. Rev. B}~{\bf 102}, 104502 (2020)}.
%
\bibitem{Sacepe2020} 
Sac\'{e}p\'{e}, B., Feigel'man, M. \& Klapwijk, T. M. Quantum breakdown of superconductivity in low-dimensional materials. \href{https://doi.org/10.1038/s41567-020-0905-x}{{\it Nat. Phys.}~{\bf 16}, 734 (2020)}.
%
\bibitem{Mooji2006}
Mooji, J. E. \& Nazarov, Y. V. 
Superconducting nanowires as quantum phase-slip junctions. \href{https://doi.org/10.1038/nphys234}{{\it Nat. Phys.}~{\bf 2}, 169 (2006)}.
%
\bibitem{deGraaf2018} 
de Graaf, S. E., Skacel, S. T., H\"{o}nigl-Decrinis, T., Shaikhaidarov, R., Rotzinger, H., Linzen, S., Ziegler, M., H\"{u}bner, U., Meyer, H.-G., Antonov, V., Il'ichev, E., Ustinov, A. V., Tzalenchuk, A. Y. \& Astafiev, O. V. Charge quantum interference device. \href{https://doi.org/10.1038/s41567-018-0097-9}{{\it Nat. Phys.}~{\bf 14}, 590 (2018)}.
%
\bibitem{Astafiev2012} 
Astafiev, O. V., Ioffe, L. B., Kafanov, S., Pashkin, Y. A., Arutyunov, K. Y., Shahar, D, Cohen, O. 
\& Tsai, J. S. Coherent quantum phase slip. \href{https://doi.org/10.1038/nature10930}{{\it Nature}~{\bf 484}, 355 (2012)}.
%
\bibitem{Lyatti2020} 
Lyatti, M., Wolff, M. A., Gundareva, I., Kruth, M., Ferrari, S., Dunin-Borkowski, R. E.  \& Schuck, C. Energy-level quantization and single-photon control of phase slips in ${\rm YBa}_2{\rm Cu}_3{\rm O}_{7-x}$ nanowires. \href{https://doi.org/10.1038/s41467-020-14548-x}{{\it Nat. Commun.}~{\bf 11}, 763 (2020)}.
%
\bibitem{Mooji2005} 
Mooji, J. E. \& Harmans, C. J. P. M. Phase-slip flux qubits. \href{https://doi.org/10.1088/1367-2630/7/1/219}{{\it New J. Phys.}~{\bf 7}, 219 (2005)}.
%
\bibitem{Place2021}
Place, A. P. M., Rodgers, L. V. H., Mundada, P., Smitham, B. M., Fitzpatrick, M.,  Leng, Z.,  Premkumar, A., Bryon, J., Vrajitoarea, A., Sussman, S., Cheng, G., Madhavan, T., Babla, H. K., Le, X. H., Gang, Y., J\"{a}ck, B., Gyenis, A., Yao, N., Cava, R. J., de Leon, N. P. \& Houck, A. A. 
New material platform for superconducting transmon qubits with coherence times exceeding $0.3$ milliseconds. \href{https://doi.org/10.1038/s41467-021-22030-5}{{\it Nat. Commun.}~{\bf 12}, 1779 (2021)}.
%
\bibitem{McRae2021}
McRae, C. R. H., McFadden, A., Zhao, R., Wang, H., Long, J. L., Zhao, T., Park, S., Bal, M.,  Palmstr{\o}m, C. J. \& Pappas, D. P. Cryogenic microwave loss in epitaxial Al/GaAs/Al trilayers for superconducting circuits. \href{https://doi.org/10.1063/5.0029855}{{\it J. Appl. Phys.}~{\bf 129}, 025109 (2021)}.
%
\bibitem{Melville2020} 
Melville, A., Calusine, G., Woods, W., Serniak, K., Golden, E., Niedzielski, M., Kim, D. K., Sevi, A., Yoder, J. L., Dauler, E. A. \& Oliver, W. D. Comparison of dielectric loss in titanium nitride and aluminum superconducting resonators. \href{https://doi.org/10.1063/5.0021950}{{\it Appl. Phys. Lett.}~{\bf 117}, 124004 (2020)}.
%
\bibitem{Kim2021} 
Kim, S., Terai, H., Yamashita, T., Qiu, W., Fuse, T., Yoshihara, F.,  Ashhab, S., Inomata, K. \& Semba, K. Enhanced coherence of all-nitride superconducting qubits epitaxially grown on silicon substrate. \href{https://doi.org/10.1038/s43246-021-00204-4}{{\it Commun. Mater.}~{\bf 2}, 98 (2021)}.
%
\bibitem{Schmidt2018} 
Schmidt, F. E., Jenkins, M. D., Watanabe, K., Taniguchi, T. \& Steele, G. A. A ballistic graphene superconducting microwave circuit. \href{https://doi.org/10.1038/s41467-018-06595-2}{{\it Nat. Commun.}~{\bf 9}, 4069 (2018)}.
%
\bibitem{Lee19} 
Lee, K.-H., Chakram, S.,  Kim, S. E.,  Mujid, F., Ray, A., Gao, H., Park, C., Zhong, Y., Muller, D. A., Schuster, D. I. \&  Park, J. Two-dimensional material tunnel barrier for Josephson junctions and superconducting qubits. \href{https://doi.org/10.1021/acs.nanolett.9b03886}{{\it Nano Lett.}~{\bf 19}, 8287 (2019)}.
%
\bibitem{Wang2019} 
Wang, J. I.-J., Rodan-Legrain, D., Bretheau, L., Campbell, D. L., Kannan, B., Kim, D., Kjaergaard, M., Krantz, P., Samach, G. O., Yan, F., Yoder, J. L., Watanabe, K., Taniguchi, T., Orlando, T. P.,  Gustavsson, S., Jarillo-Herrero, P. \& Oliver, W. D. Coherent control of a hybrid superconducting circuit made with graphene-based van der Waals heterostructures. \href{https://doi.org/10.1038/s41565-018-0329-2}{{\it Nat. Nanotech.}~{\bf 14}, 120 (2019)}.
%
\bibitem{Antony2021} 
Antony, A., Gustafsson, M. V., Ribeill, G., Ware, M., Rajendran, A., Govia, L. C. G., Ohki, T. A., Taniguchi, T., Watanabe, K., Hone, J. \& Fong, K. C. Miniaturizing transmon qubits using van der Waals materials. \href{https://arxiv.org/abs/2109.02824}{{\it arXiv:2109.02824} (2021)}.
%
\bibitem{Wang2021}
Wang, J. I-J., Yamoah, M. A., Li, Q., Karamlou, A., Dinh, T., Kannan, B., Braumueller, J., Kim, D., Melville, A. J., Muschinske, S. E., Niedzielski, B. M., Serniak, K., Sung, Y., Winik, R., Yoder, J. L., Schwartz, M., Watanabe, K., Taniguchi, T., Orlando, T. P., Gustavsson, S., Jarillo-Herrero, P. \& Oliver, W. D. Hexagonal Boron Nitride (hBN) as a low-loss dielectric for superconducting quantum circuits and qubits. \href{https://arxiv.org/abs/2109.00015}{{\it arXiv:2107.09147} (2021)}.
%
\bibitem{Rhodes2019}
Rhodes, D., Chae, S. H., Ribeiro-Palau, R. \& Hone, J. Disorder in van der Waals heterostructures of 2D materials.
\href{https://doi.org/10.1038/s41563-019-0366-8}{{\it Nat. Mater.}~{\bf 18}, 541 (2019)}.
%
\bibitem{Akinwande2019}
Akinwande, D., Huyghebaert, C., Wang, C.-H., Serna, M. I.,  Goossens, S., Li, L.-J., Philip Wong, H.-S. \& Koppens, F. H. L. Graphene and two-dimensional materials for silicon technology.
\href{https://doi.org/10.1038/s41586-019-1573-9}{{\it Nature}~{\bf 573}, 507 (2019)}.
%
\bibitem{Asselberghs2020}
Asselberghs, I., Smets, Q., Schram, T., Groven, B., Verreck, D.,  Afzalian, A., Arutchelvan, G., Gaur, A., Cott, D., Maurice, T.,  Brems, S., Kennes, K., Phommahaxay, A., Dupuy, E., Radisic, D., de Marneffe, J.-F., Thiam, A., Li, W., Devriendt, K., Huyghebaert, C.,  Lin, D., Caymax, M., Morin, P. \& Radu, I. P. Wafer-scale integration of double gated ${\rm WS}_2$-transistors in 300mm Si CMOS fab. \href{https://doi.org/10.1109/IEDM13553.2020.9371926}{{\it 2020 IEEE International Electron Devices Meeting (IEDM)}, 40.2.1 (2020)}.
%
\bibitem{Gokhale2020} 
Gokhale, V. J., Downey, B. P., Katzer, D. S., Nepal, N., Lang, A. C., Stroud, R. M. \& 
Meyer, D. J. Epitaxial bulk acoustic wave resonators as highly coherent multi-phonon sources for quantum acoustodynamics. \href{https://doi.org/10.1038/s41467-020-15472-w}{{\it Nat. Commun.}~{\bf 11}, 1 (2020)}.
%
\bibitem{Zhao2020}
Zhao, R., Park, S., Zhao, T., Bal, M., McRae, C. R. H., Long, J. \& Pappas, D. P. 
Merged-element transmon. \href{https://doi.org/10.1103/PhysRevApplied.14.064006}{{\it Phys. Rev. Applied}~{\bf 14}, 064006 (2020)}.
%
\bibitem{Polonsky1995} 
Polonsky, S. V., Kirichenko, A. F., Semenov, V. K.  \& Likharev, K. K. Rapid single flux quantum random access memory. \href{https://doi.org/10.1109/77.403223}{{\it IEEE Trans. Appl. Supercon.}~{\bf 5}, 3000 (1995)}.
%
\bibitem{Ilin2021} 
Ilin, E., Song, X., Burkova, I., Silge, A., Guo, Z., Ilin, K. \& Bezryadin, A. Supercurrent-controlled kinetic inductance superconducting memory element. \href{https://doi.org/10.1063/5.0040563}{{\it Appl. Phys. Lett.}~{\bf 118}, 112603 (2021)}.
%
\bibitem{Murphy2017} 
Murphy, A., Averin, D. V. \& Bezryadin, A. Nanoscale superconducting memory based on the kinetic inductance of asymmetric nanowire loops, \href{https://doi.org/10.1088/1367-2630/aa7331}{{\it New J. Phys.}~{\bf 19}s, 063015 (2017)}.
%
\bibitem{Ligato2021a} 
Ligato, N., Strambini, E., Paolucci, F. \& Giazotto, F. Preliminary demonstration of a persistent Josephson phase-slip memory cell with topological protection. \href{https://doi.org/10.1038/s41467-021-25209-y}{{\it Nat. Commun.}~{\bf 12}, 5200 (2021)}.
%
\bibitem{Madden2018} 
Madden, A. E., Willard, J. C., Loloee, R. \& Birge, N. O. Phase controllable Josephson junctions for cryogenic memory. \href{https://doi.org/10.1088/1361-6668/aae8cf}{{\it Supercond. Sci. Technol.}~{\bf 32}, 015001 (2019)}.
%
\bibitem{Karelina2021} 
Karelina, L. N., Hovhannisyan, R. A., Golovchanskiy, I. A., Chichkov, V. I., Ben Hamida, A.,  Stolyarov, V. S., Uspenskaya, L. S., Erkenov, Sh. A., Bolginov, V. V. \& Ryazanov, V. V. Scalable memory elements based on rectangular SIsFS junctions. \href{https://doi.org/10.1063/5.0063274}{{\it J. Appl. Phys.}~{\bf 130}, 173901 (2021)}.
%
\bibitem{Baek2014}
Baek, B., Rippard, W. H., Benz, S. P., Russek, S. E. \& Dresselhaus, P. D. Hybrid superconducting-magnetic memory device using competing order parameters  \href{https://doi.org/10.1038/ncomms4888}{{\it Nat. Commun.}~{\bf 5}, 3888 (2014)}.
%
\bibitem{Ligato2021b} 
Ligato, N., Paolucci, F., Strambini, E. \& Giazotto, F. Thermal superconducting quantum interference proximity transistor. \href{https://arxiv.org/abs/2107.08936}{{\it arXiv:2107.08936v1} (2021)}.
%
\bibitem{Fornieri2017} 
Fornieri, A. \& Giazotto, F. Towards phase-coherent caloritronics in superconducting circuits. \href{https://doi.org/10.1038/nnano.2017.204}{{\it Nat. Nanotech.}~{\bf 12}, 944 (2017)}.
%
\bibitem{McArdle2020}
McArdle, S., Endo, S., Aspuru-Guzik, A., Benjamin, S. C. 
\& Yuan, X. Quantum computational chemistry. \href{https://doi.org/10.1103/RevModPhys.92.015003}{{\it Rev. Mod. Phys.}~{\bf 92}, 015003 (2020)}.
%
\bibitem{Jordan1928}
Jordan, P. \& Wigner, E. \"{U}ber das Paulische \"{a}quivalenzverbot. \href{https://doi.org/10.1007/BF01331938}{{\it Z. Phys.}~{\bf 47}, 631 (1928)}.
%
\bibitem{Bravyi2002}
Bravyi, S. B. \& Kitaev, A. Y. Fermionic quantum computation. \href{https://doi.org/10.1006/aphy.2002.6254}{{\it Ann. Phys.}~{\bf 298}, 210 (2002)}.
%
\bibitem{Peruzzo2014}
Peruzzo, A., McClean, J., Shadbolt, P., Yung, M.-H., Zhou, X.-Q., Love, P. J., Aspuru-Guzik, A. \& O'Brien, J. L. A variational eigenvalue solver on a photonic quantum processor. \href{https://doi.org/10.1038/ncomms5213}{{\it Nat. Commun.}~{\bf 5}, 4213 (2014)}.
%
\bibitem{Kitaev1995}
Kitaev, A. Y. Quantum measurements and the Abelian stabilizer problem. \href{https://arxiv.org/abs/quant-ph/9511026}{{\it arXiv:quant-ph/9511026} (1995)}.
%
\bibitem{Abrams1999}
Abrams, D. S.  \& Lloyd, S. Quantum algorithm providing exponential speed increase for finding eigenvalues and eigenvectors. \href{https://doi.org/10.1103/PhysRevLett.83.5162}{{\it Phys. Rev. Lett.}~{\bf 83}, 5162 (1999)}.
%
\bibitem{Preskill2018}
Preskill, J. \href{https://doi.org/10.22331/q-2018-08-06-79}{{\it Quantum}~{\bf 2}, 79 (2018)}.
%
\bibitem{Arute2020}
Arute, F., Arya, K., Babbush, R., Bacon, D., Bardin, J. C., Barends, R., Boixo, S.,  Broughton, M., Buckley, B. B., Buell, D. A., Burkett, B., Bushnell, N., Chen, Y., Chen, Z., Chiaro, B., Collins, R., Courtney, W., Demura, S., Dunsworth, A., Eppens, D., Farhi, E., Fowler, A., Foxen, B., Gidney, C., Giustina, M., Graff, R., Habegger, S.,  Harrigan, M. P., Ho, A., Hong, S., Huang, T.,  Huggins, W. J., Ioffe, L., Isakov, S. V., Jeffrey, E., Jiang, Z., Jones, C., Kafri, D., Kechedzhi, K.,  Kelly, J., Kim, S., Klimov, P. V., Korotkov, A., Kostritsa, F., Landhuis, D., Laptev, P., Lindmark, M., Lucero, E., Martin, o., Martinis, J. M., McClean, J. R., McEwen, M., Megrant, A., Mi, X., Mohseni, M.,  Mruczkiewicz, W., Mutus, J., Naaman, O., Neeley, M., Neill, C., Neven, H., Yuezhen Niu, M., O'Brien, T. E., Ostby, E., Petukhov, A., Putterman, H., Quintana, C., Roushan, P., Rubin, N. C., Sank, D., Satzinger, K. J., Smelyanskiy, V., Strain, D., Sung, K. J.,  Szalay, M., Takeshita, T. Y., Vainsencher, A., White, T., Wiebe, N., Jamie Yao, Z., Yeh, P. \& Zalcman, A. Hartree-Fock on a superconducting qubit quantum computer. \href{https://doi.org/10.1126/science.abb9811}{{\it Science}~{\bf 369}, 1084 (2020)}.
%
\bibitem{Ippoliti2021}
Ippoliti, M., Kechedzhi, K., Moessner, R., Sondhi, S.  L. \& Khemani,  V. Many-Body physics in the NISQ era: Quantum programming a discrete time crystal. \href{https://doi.org/10.1103/PRXQuantum.2.030346}{{\it PRX Quantum}~{\bf 2}, 030346 (2021)}.
%
\bibitem{Else2020}
Else, D. V., Monroe, C., Nayak, C. \& Yao, N. Y. Discrete time crystals. \href{https://doi.org/10.1146/annurev-conmatphys-031119-050658}{{\it Annu. Rev. Condens. Matter Phys.}~{\bf 11}, 467 (2020)}.
%
\bibitem{Rossini2021}
Rossini, D. \& Vicari, E. Coherent and dissipative dynamics at quantum phase transitions. \href{https://doi.org/10.1016/j.physrep.2021.08.003}{{\it Phys. Rep.}~{\bf 936}, 1 (2021)}.
%
\bibitem{Dupont2021}
Dupont, M. \& Moore, J. E. Quantum criticality using a superconducting quantum processor. \href{https://arxiv.org/abs/2109.10909}{{\it arXiv:2109.10909} (2021)}.
%
\bibitem{Troyer2005}
Troyer, M. \& Wiese, U.-J.  Computational complexity and fundamental limitations to fermionic quantum Monte Carlo simulations. \href{https://doi.org/10.1103/PhysRevLett.94.170201}{{\it Phys. Rev. Lett.}~{\bf 94}, 170201 (2005)}.
%
\bibitem{Banuls2020}
Ba\~{n}uls, M. C., Blatt, R., Catani, J., Celi, A., Cirac, J. I., Dalmonte, M., Fallani, L.,  Jansen, K.,  Lewenstein, M., Montangero, S., Muschik, C. A., Reznik, B., Rico, E.,  Tagliacozzo, L., Van Acoleyen, K., Verstraete, F.,  Wiese, U.-J., Wingate, M., Zakrzewski, J. \& Zoller, P. Simulating lattice gauge theories within quantum technologies. \href{https://doi.org/10.1140/epjd/e2020-100571-8}{{\it Eur. Phys. J. D}~{\bf 74}, 165 (2020)}.
%
\bibitem{Kemme2011}
Temme, K., Osborne, T. J., Vollbrecht, K. G., Poulin, D. \& Verstraete, F.  Quantum Metropolis sampling. \href{https://doi.org/10.1038/nature09770}{{\it Nature}~{\bf 471}, 87 (2011)}.
%
\bibitem{Clemente2020} 
Clemente, G., Cardinali, M., Bonati, C., Calore, E., Cosmai, L., D'Elia, M., Gabbana, A., Rossini, D., Schifano, F. S., Tripiccione, R. \& Vadacchino, D. Quantum computation of thermal averages in the presence of a sign problem. \href{https://doi.org/10.1103/PhysRevD.101.074510}{{\it Phys. Rev. D}~{\bf 101}, 074510 (2020)}.
%
\bibitem{Yung2012}
Yung, M.-H. \& Aspuru-Guzik, A. A quantum-quantum Metropolis algorithm. \href{https://doi.org/10.1073/pnas.1111758109v}{{\it Proc. Natl. Acad. Sci. (USA)}~{\bf 109}, 754 (2012)}.
%
\bibitem{Motta2020}
Motta, M., Sun, C., Tan, A. T. K., O'Rourke, M. J., Ye, E., Minnich, A. J., Brand\~{a}o, F. G. S. L., \&  Chan, G. K.-L. Determining eigenstates and thermal states on a quantum computer using quantum imaginary time evolution. \href{https://doi.org/10.1038/s41567-019-0704-4}{{\it Nat. Phys.}~{\bf 16}, 205 (2020)}.
%
\bibitem{Boccali2019}
Boccali, T. Computing models in high energy physics. \href{https://doi.org/10.1016/j.revip.2019.100034}{{\it Rev. Phys.}~{\bf 4}, 100034 (2019)}.
%
\bibitem{Schmidt2016}
Schmidt, B. The high-luminosity upgrade of the LHC: Physics and technology challenges for the accelerator and the experiments. \href{https://doi.org/10.1088/1742-6596/706/2/022002}{{\it J. Phys.: Conf. Ser.}~{\bf 706}, 022002 (2016)}.
%
\bibitem{CMS}
The CMS Collaboration, ``Evolution of the CMS Computing Model towards Phase-2'', \href{https://cds.cern.ch/record/2751565?ln=it}{CERN-CMS-NOTE-2021-001}. 
%
\bibitem{Bapst2020}
Bapst, F.,  Bhimji, W.,  Calafiura, P., Gray, H., Lavrijsen, W., Linder, L. \& Smith, A. A pattern recognition algorithm for quantum annealers. \href{https://doi.org/10.1007/s41781-019-0032-5}{{\it Comput Softw. Big Sci.}~{\bf 4}, 1 (2020)}.
%
\bibitem{Tuysuz2020}
T\"{u}ys\"{u}z, C., Carminati, F., Demirk\"{o}z, B., Dobos, D., Fracas, F., Novotny, K., Potamianos, K., Vallecorsa, S. \& Vlimant, J.-R. Particle track reconstruction with quantum algorithms. \href{https://doi.org/10.1051/epjconf/202024509013}{{\it EPJ Web Conf.}~{\bf 245}, 09013 (2020)}.
%
\bibitem{Mott2017}
Mott, A., Job, J., Vlimant, J.-R., Lidar, D. \& Spiropulu, M. Solving a Higgs optimization problem with quantum annealing for machine learning. \href{https://doi.org/10.1038/nature24047}{{\it Nature}~{\bf 550}, 375 (2017)}.
%
\bibitem{Felser2021}
Felser, T., Trenti, M., Sestini, L., Gianelle, A., Zuliani, D., Lucchesi, D. \& Montangero, S. Quantum-inspired machine learning on high-energy physics data. \href{https://doi.org/10.1038/s41534-021-00443-w}{{\it npj Quant. Inf.}~{\bf 7}, 111 (2021)}.
%
\bibitem{Wu2021}
Wu, S. L., Sun, S., Guan, W., Zhou, C., J. Chan, C. L. Cheng, T. Pham, Y. Qian, A. Z. Wang, R. Zhang, M. Livny, J. Glick, P. Kl. Barkoutsos, S. Woerner, I. Tavernelli, F. Carminati, A. Di Meglio, A. C. Y. Li, J. Lykken, P. Spentzouris, S. Y.-C. Chen, S. Yoo, and T.-C. Wei, Application of quantum machine learning using the quantum kernel algorithm on high energy physics analysis at the LHC. \href{https://doi.org/10.1103/PhysRevResearch.3.033221}{{\it Phys. Rev. Res.}~{\bf 3}, 033221 (2021)}.
%
\bibitem{Bauer2021}
Bauer, C. W., de Jong, W. A., Nachman, B. \& Provasoli, D. A quantum algorithm for high energy physics simulations. \href{https://doi.org/10.1103/PhysRevLett.126.062001}{{\it Phys. Rev. Lett.}~{\bf 126}, 062001 (2021)}.
%
\bibitem{Guan2021}
Guan, W., Perdue, G., Pesah, A., Schuld, M., Terashi, K., Vallecorsa, S. \& Vlimant, J.-R. Quantum machine learning in high energy physics. \href{https://doi.org/10.1088/2632-2153/abc17d}{{\it Mach. Learn.: Sci. Technol.}~{\bf 2}, 011003 (2021)}.
%
\bibitem{Bennett1998}
Bennett, C. H. \& Shor, P. W. Quantum information theory. 
\href{https://doi.org/10.1109/18.720553}{{\it IEEE Trans. Inf. Theory}~{\bf 44}, 2724 (1998)}.
%
\bibitem{Bennett1993}
Bennett, C. H., Brassard, G., Cr\'{e}peau, C., Jozsa, R., Peres, A. \& Wootters, W. K. Teleporting an unknown quantum state via dual classical and Einstein-Podolsky-Rosen channels. \href{https://doi.org/10.1103/PhysRevLett.70.1895}{{\it Phys. Rev. Lett.}~{\bf 70}, 1895 (1993)}.
%
\bibitem{Wootters1982}
Wootters, W. \& Zurek, W. A single quantum cannot be cloned. \href{https://doi.org/10.1038/299802a0}{{\it Nature}~{\bf 299}, 802 (1982)}.
%
\bibitem{Bennett1992}
Bennett, C. H. \& Wiesner, S. J. Communication via one- and two-particle operators on Einstein-Podolsky-Rosen states. \href{https://doi.org/10.1103/PhysRevLett.69.2881}{{\it Phys. Rev. Lett.}~{\bf 69}, 2881 (1992)}.
%
\bibitem{Bennett2014}
Bennett, C. H. \& Brassard, G. Quantum cryptography: Public key distribution and coin tossing. \href{https://doi.org/10.1016/j.tcs.2014.05.025}{{\it Theor. Comput. Sci.}~{\bf 560}, 7 (2014)}.
%
\bibitem{Gisin2002}
Gisin, N., Ribordy, G., Tittel, W. \& Zbinden, H. 
Quantum cryptography. \href{https://doi.org/10.1103/RevModPhys.74.145}{{\it Rev. Mod. Phys.}~{\bf 74}, 145 (2002)}.
%
\bibitem{Lu2019a}
Lu, X., Li, Q., Westly, D. A., Moille, G., Singh, A., Anant, V. \& Srinivasan, K. Chip-integrated visible-telecom entangled photon pair source for quantum communication. \href{https://doi.org/10.1038/s41567-018-0394-3}{{\it Nat. Phys.}~{\bf 15}, 373 (2019)}.
%
\bibitem{Sellenart2017} 
Senellart, P., Solomon, G. \& White, A. High-performance semiconductor quantum-dot single-photon sources. \href{https://doi.org/10.1038/nnano.2017.218}{{\it Nat. Nanotech.}~{\bf 12}, 1026 (2017)}.
%
\bibitem{Tomm2021} 
Tomm, N., Javadi, A., Antoniadis, N. O., Najer, D., L\"{o}bl, M. C., Korsch, A. R., Schott, R., Valentin, S. R., Wieck, A. D., Ludwig, A. \& Warburton, R. J. A bright and fast source of coherent single photons. \href{https://doi.org/10.1038/s41565-020-00831-x}{{\it Nat. Nanotech.}~{\bf 16}, 399 (2021)}.
%
%
\bibitem{Fedyanin2021}
Fedyanin, D. Y. Optoelectronics of color centers in diamond and silicon carbide: From single-photon luminescence to electrically controlled spin qubits. \href{https://doi.org/10.1002/qute.202100048}{{\it Adv. Quantum Technol.}~{\bf 4}, 2100048 (2021)}.
%
\bibitem{Grosso2017}
Grosso, G., Moon, H., Lienhard, B., Ali, S., Efetov, D. K., Furchi, M. M., Jarillo-Herrero, P., Ford, M. J., Aharonovich, I. \& Englund, D. Tunable and high-purity room temperature single-photon emission from atomic defects in hexagonal boron nitride. \href{https://doi.org/10.1038/s41467-017-00810-2}{{\it Nat. Commun.}~{\bf 8}, 705 (2017)}.
%
\bibitem{Schuler2020}
Schuler, B., Cochrane, K. A., Kastl, C., Barnard, E. S., Wong, E., Borys, N. J., Schwartzberg, A. M., Ogletree, D. F., de Abajo, F. J. G., Weber-Bargioni, A. Electrically driven photon emission from individual atomic defects in monolayer ${\rm WS}_2$. \href{https://doi.org/10.1126/sciadv.abb5988}{{\it Sci. Adv.}~{\bf 6}, eabb5988 (2020)}.
%
\bibitem{Romagnoli2018}
Romagnoli, M., Sorianello, V., Midrio, M., Koppens, F. H. L., Huyghebaert, C., Neumaier, D., Galli, P.,  Templ, W., D’Errico, A. \& Ferrari, A. C. Graphene-based integrated photonics for next-generation datacom and telecom. \href{https://doi.org/10.1038/s41578-018-0040-9}{{\it Nat. Rev. Mater.}~{\bf 3}, 392 (2018)}.
%
\bibitem{Turunen2022}
Turunen, M., Brotons-Gisbert, M., Dai, Y., Wang, Y., Scerri, E., Bonato, C., J\"{o}ns, K. D., Sun, Z. \& Gerardot, B. D. Quantum photonics with layered 2D materials. \href{https://doi.org/10.1038/s42254-021-00408-0}{{\it Nat. Rev. Phys.} (2022)}.
%
\bibitem{Hadfield2009}
Hadfield, R. H.  Single-photon detectors for optical quantum information applications. \href{https://doi.org/10.1038/nphoton.2009.230}{{\it Nat. Photon.}~{\bf 3}, 696 (2009)}.
%
\bibitem{Vallone2015}
Vallone, G., Bacco, D., Dequal, D., Gaiarin, S., Luceri, V., Bianco, G. \&   Villoresi, P. Experimental satellite quantum communications. \href{https://doi.org/10.1103/PhysRevLett.115.040502}{{\it Phys. Rev. Lett.}~{\bf 115}, 040502 (2015)}.
%
\bibitem{Liao2017}
Liao, S.-K, Cai, W.-Q., Liu, W.-Y., Zhang, L., Li, Y., Ren, J.-G., Yin, J., Shen, Q., Cao, Y., Li, Z.-P., Li, F.-Z., Chen, X.-W., Sun, L.-H., Jia, J.-J., Wu, J.-C., Jiang, X.-J., Wang, J.-F., Huang, Y.-M., Wang, Q., Zhou, Y.-L., Deng, L., Xi, T., Ma, L., Hu, T., Zhang, Q., Chen, Y.-A., Liu, N.-L., Wang, X.-B., Zhu, Z.-C., Lu, C.-Y., Shu, R., Peng, C.-Z., Wang, J.-Y. \& Pan, J.-W. Satellite-to-ground quantum key distribution. \href{https://doi.org/10.1038/nature23655}{{\it Nature}~{\bf 549}, 43 (2017)}.
%
\bibitem{Korzh2020} 
Korzh, B., Zhao, Q.-Y., Allmaras, J. P., Frasca, S., Autry, T. M., Bersin, E. A., Beyer, A. D., Briggs, R. M., Bumble, B., Colangelo, M., Crouch, G. M., Dane, A. E., Gerrits, T., Lita, A. E., Marsili, F., Moody, G., Pe\~{n}a, C., Ramirez, E., Rezac, J. D., Sinclair, N., Stevens, M. J., Velasco, A. E., Verma, V. B., Wollman, E. E., Xie, S., Zhu, D., Hale, P. D., Spiropulu, M., Silverman, K. L., Mirin, R. P., Nam, S. W., Kozorezov, A. G., Shaw, M. D. \& Berggren, K. K. Demonstration of sub-$3~{\rm ps}$ temporal resolution with a superconducting nanowire single-photon detector. \href{https://doi.org/10.1038/s41566-020-0589-x}{{\it Nat. Photon.}~{\bf 14}, 250 (2020)}.
%
\bibitem{Zadeh2021} 
Zadeh, I. E., Chang, J., Los, J. W. N., Gyger, S., Elshaari, A. W., Steinhauer, S., Dorenbos, S. N. \& Zwiller, V. Superconducting nanowire single-photon detectors: A perspective on evolution, state-of-the-art, future developments, and applications. \href{https://doi.org/10.1063/5.0045990}{{\it Appl. Phys. Lett.}~{\bf 118}, 190502 (2021)}.
%
\bibitem{Prabhakar2020}
Prabhakar, S., Shields, T., Dada, A. C., Ebrahim, M., Taylor, G. G., Morozov, D., Erotokritou, K., 
Miki, S., Yabuno, M., Terai, H., Gawith, C., Kues, M., Caspani, L., Hadfield, R. H. \& Clerici, M. Two-photon quantum interference and entanglement at $2.1~{\rm \mu m}$. \href{https://doi.org/10.1126/sciadv.aay5195}{{\it Sci. Adv.}~{\bf 6}, eaay5195 (2020)}.
%
\bibitem{Sibson2017}
Sibson, P., Erven, C., Godfrey, M., Miki, S., Yamashita, T., Fujiwara, M., Sasaki, M., Terai, H.,  Tanner, M. G., Natarajan, C. M., Hadfield, R. H., O'Brien, J. L. \& Thompson, M. G. Chip-based quantum key distribution. \href{https://doi.org/10.1038/ncomms13984}{{\it Nat. Commun.}~{\bf 8}, 13984 (2017)}.
%
\bibitem{Elshaari2020}
Elshaari, A. W., Pernice, W., Srinivasan, K., Benson, O. \& Zwiller, V. Hybrid integrated quantum photonic circuits. \href{https://doi.org/10.1038/s41566-020-0609-x}{{\it Nat. Photon.}~{\bf 14}, 285 (2020)}.
%
\bibitem{Paraiso2021}
Para\"{i}so, T. K., Roger, T., Marangon, D. G., De Marco, I., Sanzaro, M., Woodward, R. I., Dynes, J. F., Yuan, Z. \& Shields, A. J. A photonic integrated quantum secure communication system.
\href{https://doi.org/10.1038/s41566-021-00873-0}{{\it Nat. Photon.}~{\bf 15}, 850 (2021)}.
%
\bibitem{Beutel2021}
Beutel, F., Gehring, H., Wolff, M. A., Schuck, C. \& Pernice, W. Detector-integrated on-chip QKD receiver for GHz clock rates. \href{https://doi.org/10.1038/s41534-021-00373-7}{{\it npj Quantum Inf.}~{\bf 7}, 40 (2021)}. 
%
\bibitem{Pelucchi2021}
Pelucchi, E., Fagas, G., Aharonovich, I., Englund, D.,  Figueroa, E., Gong, Q., Hannes, H., Liu, J., Lu, C.-Y.,  Matsuda, N., Pan, J.-W., Schreck, F., Sciarrino, F.,  Silberhorn, C., Wang, J. \& J\"{o}ns, K. D. The potential and global outlook of integrated photonics for quantum technologies. \href{https://doi.org/10.1038/s42254-021-00398-z}{{\it Nat. Rev. Phys.} (2021)}. 
%
\bibitem{Rivest1978}
Rivest, R. L., Shamir, A. \& Adleman, L. A method for obtaining digital signatures and public-key cryptosystems. \href{https://doi.org/10.1145/359340.359342}{{\it Commun. of the ACM}~{\bf 21}, 120 (1978)}.
%
\bibitem{Shor1999}
Shor, P. W. Polynomial-time algorithms for prime factorization and discrete logarithms on a quantum computer. \href{https://doi.org/10.1137/S0036144598347011}{{\it SIAM Rev.}~{\bf 41}, 303 (1999)}.
%
\bibitem{Degen2017} 
Degen, C. L., Reinhard, F. \& Cappellaro, P. Quantum sensing. \href{https://doi.org/10.1103/RevModPhys.89.035002}{{\it Rev. Mod. Phys.}~{\bf 89}, 035002 (2017)}.
%
\bibitem{Bucher2019}
Bucher, D. B., Aude Craik, D. P. L., Backlund, M. P., Turner, M. J., Dor, O. B., Glenn, D. R. \& Walsworth, R. L. Quantum diamond spectrometer for nanoscale NMR and ESR spectroscopy.
\href{https://doi.org/10.1038/s41596-019-0201-3}{{\it Nat. Protoc.}~{\bf 14}, 2707 (2019)}.
%
\bibitem{Casola2018}
Casola, F., van der Sar, T. \& Yacoby, A. Probing condensed matter physics with magnetometry based on nitrogen-vacancy centres in diamond. \href{https://doi.org/10.1038/natrevmats.2017.88}{{\it Nat. Rev. Mater.}~{\bf 3}, 17088 (2018)}. 
%
\bibitem{Yang2020}
Yang, B., Murooka, T., Mizuno, K., Kim, K., Kato, H., Makino, T., Ogura, M., Yamasaki, S., Schmidt, M. E., Mizuta, H., Yacoby, A., Hatano, M. \& Iwasaki, T. Vector electrometry in a wide-gap-semiconductor device using a spin-ensemble quantum sensor. \href{https://doi.org/10.1103/PhysRevApplied.14.044049}{{\it Phys. Rev. Applied}~{\bf 14}, 044049 (2020)}.
%
\bibitem{Hart2021}
Hart, C. A., Schloss, J. M., Turner, M. J., Scheidegger, P. J., Bauch, E. \& Walsworth, R. L. N-V-Diamond magnetic microscopy using a double quantum $4$-Ramsey protocol. \href{https://doi.org/10.1103/PhysRevApplied.15.044020}{{\it Phys. Rev. Applied}~{\bf 15}, 044020 (2021)}.
%
\bibitem{Marsh2019} 
Marsh, D. J. E., Fong, K. C., Lentz, E. W., \v{S}mejkal, L. \& Ali, M. N. A proposal to detect dark matter using axionic topological antiferromagnets. \href{https://doi.org/10.1103/PhysRevLett.123.121601}{{\it Phys. Rev. Lett.}~{\bf 123}, 121601 (2019)}.
%
\bibitem{Bartram2021}
Bartram, C., Braine, T., Cervantes, R., Crisosto, N., Du, N., Leum, G., Mohapatra, P., Nitta, T., Rosenberg, L. J., Rybka, G., Yang, J., Clarke, J., Siddiqi, I., Agrawal, A., Dixit, A. V., Awida, M. H.,  Chou, A. S., Hollister, M., Knirck, S., Sonnenschein, A., Wester, W., Gleason, J. R., Hipp, A. T., Jois, S., Sikivie, P., Sullivan, N. S., Tanner, D. B., Hoof, S., Lentz, E., Khatiwada, R., Carosi, G., Cisneros, C., Robertson, N., Woollett, N., Duffy, L. D., Boutan, C., Jones, M., LaRoque, B. H., Oblath, N. S., Taubman, M. S., Daw, E. J., Perry, M. G., Buckley, J. H., Gaikwad, C, Hoffman, J., Murch, K., Goryachev, M, McAllister, B. T., Quiskamp, A., Thomson, C. \& Tobar, M. E. Dark matter axion search using a Josephson traveling wave parametric amplifier. \href{https://arxiv.org/abs/2110.10262}{{\it arXiv:2110.10262} (2021)}.
%
\bibitem{Lehnert2021}
Lehnert, K. W. Quantum enhanced metrology in the search for fundamental physical phenomena. \href{https://arxiv.org/abs/2110.04912}{{\it arXiv:2110.04912} (2021)}.
%
\bibitem{Dixit2021} 
Dixit, A. V., Chakram, S., He, K., Agrawal, A., Naik, R. K., Schuster, D. I. \& Chou, A. Searching for dark matter with a superconducting qubit. \href{https://doi.org/10.1103/PhysRevLett.126.141302}{{\it Phys. Rev. Lett.}~{\bf 126}, 141302 (2021)}.
%
\bibitem{Chen2021}
Chen, Y.-A., Zhang, Q., Chen, T.-Y., Cai, W.-Q., Liao, S.-K., Zhang, J., Chen, K., Yin, J., Ren, J.-G.,  Chen, Z., Han, S.-L., Yu, Q., Liang, K., Zhou, F., Yuan, X., Zhao, M.-S., Wang, T.-Y., Jiang, X.,  Zhang, L., Liu, W.-Y., Li, Y., Shen, Q., Cao, Y., Lu, C.-Y., Shu, R., Wang, J.-Y., Li, L., Liu, N.-L., Xu, F., Wang, X.-B., Peng, C.-Z. \& Pan, J.-W. An integrated space-to-ground quantum communication network over 4,600 kilometres. \href{https://doi.org/10.1038/s41586-020-03093-8}{{\it Nature}~{\bf 589}, 214 (2021)}.
%
\bibitem{Mazin2021} 
Mazin, B. A. Superconducting materials for microwave kinetic inductance detectors. \href{https://arxiv.org/abs/2004.14576}{{\it arXiv:2004.14576} (2020)}.
%
\bibitem{Frasca2019} 
Frasca, S., Korzh, B., Colangelo, M., Zhu, D., Lita, A. E., Allmaras, J. P., Wollman, E. E., Verma, V. B., Dane, A. E., Ramirez, E., Beyer, A. D., Nam, S. W., Kozorezov, A. G., Shaw, M. D. \& Berggren, K. K. Determining the depairing current in superconducting nanowire single-photon detectors. \href{https://doi.org/10.1103/PhysRevB.100.054520}{{\it Phys. Rev. B}~{\bf 100}, 054520 (2019)}.
%
\bibitem{Zhao2020b} 
Zhao, S., Withington, S., Goldie, D. J. \& Thomas, C. N. Suppressed-gap millimetre wave kinetic inductance detectors using DC-bias current. \href{https://doi.org/10.1088/1361-6463/ab8d6d}{{\it J. Phys. D: Appl. Phys.}~{\bf 53}, 345301 (2020)}.
%
\bibitem{Shibata2010} 
Shibata, H., Takesue, H., Honjo, T., Akazaki, T. \& Tokura, Y. Single-photon detection using magnesium diboride superconducting nanowires. \href{https://doi.org/10.1063/1.3518723}{{\it Appl. Phys. Lett.}~{\bf 97}, 212504 (2010)}.
%
\bibitem{Liu2019}
Liu, X. \& Hersam, M. C. 2D materials for quantum information science. 
\href{https://doi.org/10.1038/s41578-019-0136-x}{{\it Nat. Rev. Mater.}~{\bf 4}, 669 (2019)}.
%
\bibitem{Konstantatos2012}
Konstantatos, G., Badioli, M., Gaudreau, L.,  Osmond, J., Bernechea, M., Pelayo Garcia de Arquer, F., Gatti, F. \& Koppens, F. H. L.  Hybrid graphene–quantum dot phototransistors with ultrahigh gain. \href{https://doi.org/10.1038/nnano.2012.60}{{\it Nat. Nanotech.}~{\bf 7}, 363 (2012)}.
%
\bibitem{Walsh2021} 
Walsh, E. D., Jung, W., Lee, G.-H., Efetov, D. K., Wu, B.-I., Huang, K.-F., Ohki, T. A., Taniguchi, T., Watanabe, K., Kim, P., Englund, D. \& Fong, K. C. Josephson-junction infrared single-photon detector. \href{https://doi.org/10.1126/science.abf5539}{{\it Science}~{\bf 372}, 409 (2021)}.
%
\bibitem{Lee2020} 
Lee, G.-H., Efetov, D. K., Jung, W., Ranzani, L., Walsh, E. D., Ohki, T. A., Taniguchi, T., Watanabe, K., Kim,  P., Englund, D. \& Fong, K. C. Graphene-based Josephson junction microwave bolometer. \href{https://doi.org/10.1038/s41586-020-2752-4}{{\it Nature}~{\bf 586}, 42 (2020)}.
%
\bibitem{Kokkoniemi2020} 
Kokkoniemi, R., Girard, J.-P., Hazra, D., Laitinen, A., Govenius, J., Lake, R. E., Sallinen, I., Vesterinen, V., Partanen, M., Tan, J. Y., Chan, K. W., Tan, K. Y., Hakonen, P. \& M\"{o}tt\"{o}nen, M. Bolometer operating at the threshold for circuit quantum electrodynamics. \href{https://doi.org/10.1038/s41586-020-2753-3}{{\it Nature}~{\bf 586}, 47 (2020)}.
%
\bibitem{Seifert2020} 
Seifert, P., Lu, X., Stepanov, P., Retamal, J. R. D., Moore, J. N., Fong, K.-C., Principi, A. 
\& Efetov, D. K. Magic-angle bilayer graphene nanocalorimeters: toward broadband, energy-resolving single photon detection. \href{https://doi.org/10.1021/acs.nanolett.0c00373}{{\it Nano Lett.}~{\bf 20}, 3459 (2020)}.
%
\bibitem{Checchin2021}
Checchin, M., Frolov, D., Lunin, A., Grassellino, A. \& Romanenko,  A. Measurement of the low-temperature loss tangent of high-resistivity silicon wafers with high-Q superconducting resonators. \href{https://arxiv.org/abs/2108.08894}{{\it arXiv:2108.08894} (2021)}.
%
\bibitem{Peccei1977}
Peccei, R. D. \& Quinn, H. R. CP conservation in the presence of pseudoparticles. \href{https://doi.org/10.1103/PhysRevLett.38.1440}{{\it Phys. Rev. Lett.}~{\bf 38}, 1440 (1977)}.
%
\bibitem{Hanneke2008} 
Hanneke, D., Fogwell, S. \& Gabrielse, G. New measurement of the electron magnetic moment and the fine structure constant. \href{https://doi.org/10.1103/PhysRevLett.100.120801}{{\it Phys. Rev. Lett.}~{\bf 100}, 120801 (2008)}.
%
\bibitem{Golm2021} 
Golm, J., Arguedas Cuendis, S., Calatroni, S., Cogollos, C., D\"{o}brich, B., Gallego, J. D., Garc\'{i}a Barcel\'{o}, J. M., Granados, X., Gutierrez, J., Irastorza, I. G., Koettig, T., Lamas, N., Liberadzka-Porret, J., Malbrunot, C., Millar, W. L., Navarro, P., Pereira Carlos, C., Puig, T., Rosaz, G. J., Siodlaczek, M., Telles, G. \& Wuensch, W. Thin film (high temperature) superconducting radiofrequency cavities for the search of axion dark matter. \href{https://arxiv.org/abs/2110.01296}{{\it arXiv:2110.01296} (2021)}.
%
\bibitem{Alesini2021} 
Alesini, D., Braggio, C., Carugno, G., Crescini, N., D' Agostino, D., Di Gioacchino, D., Di Vora, R.,  Falferi, P., Gambardella, U., Gatti, C., Iannone, G., Ligi, C., Lombardi, A., Maccarrone, G., Ortolan, A., Pengo, R., Pira, C., Rettaroli, A., Ruoso, G., Taffarello, L. \& Tocci, S. Realization of a high quality factor resonator with hollow dielectric cylinders for axion searches. \href{https://doi.org/10.1016/j.nima.2020.164641}{{\it Nucl. Instrum. Methods Phys. Res. A: Accel. Spectrom. Detect. Assoc. Equip.}~{\bf 985}, 164641 (2021)}.
%
\bibitem{Virtanen2018} 
Virtanen, P., Ronzani, A. \& Giazotto, F. Josephson photodetectors via temperature-to-phase conversion. \href{https://doi.org/10.1103/PhysRevApplied.9.054027}{{\it Phys. Rev. Appl.}~{\bf 9}, 054027 (2018)}.
%
\bibitem{Heikkila2018} 
Heikkil\"{a}, T. T., Ojaj\"{a}rvi, R., Maasilta, I. J., Strambini, E., Giazotto, F. \& Bergeret, F. S.  Thermoelectric radiation detector based on superconductor-ferromagnet systems. 
\href{https://doi.org/10.1103/PhysRevApplied.10.034053}{{\it Phys. Rev. Applied}~{\bf 10}, 034053 (2018)}.
%
\bibitem{Guarcello2019}
Guarcello, C., Braggio, A., Solinas, P., Pepe, G. P. \& Giazotto, F. Josephson-threshold calorimeter. \href{https://doi.org/10.1103/PhysRevApplied.11.054074}{{\it Phys. Rev. Applied}~{\bf 11}, 054074 (2019)}.
%
\bibitem{Echternach2018} 
Echternach, P. M., Pepper, B. J., Reck, T. \& Bradford, C. M. Single photon detection of $1.5~{\rm THz}$ radiation with the quantum capacitance detector. 
\href{https://doi.org/10.1038/s41550-017-0294-y}{{\it Nat. Astron.}~{\bf 2}, 90 (2018)}.
%
\bibitem{Paolucci2020} 
Paolucci, F., Ligato, N., Buccheri, V., Germanese, G., Virtanen, P. \& Giazotto, F. Hypersensitive tunable Josephson escape sensor for gigahertz astronomy. \href{https://doi.org/10.1103/PhysRevApplied.14.034055}{{\it Phys. Rev. Applied}~{\bf 14}, 034055 (2020)}.
%
\bibitem{Altman2021} 
 Altman, E., Brown, K. R., Carleo, G., Carr, L. D.,  Demler, E.,  Chin, C.,  DeMarco, B., Economou, S. E.,  Eriksson, M. A., Fu, K.-M. C.,  Greiner, M.,  Hazzard, K. R. A.,  Hulet, R. G.,  
 Koll\'{a}r, A. J., Lev, B. L.,  Lukin, M. D.,  Ma, R.,  Mi, X., Misra, S.,  Monroe, C.,  Murch, K.,  Nazario, Z.,  Ni, K.-K.,  Potter, A. C.,  Roushan, P.,  Saffman, M.,  Schleier-Smith, M.,  Siddiqi, I.,  Simmonds, R.,  Singh, M.,  Spielman, I. B.,  Temme, K.,  Weiss, D. S.,  Vu\v{c}kovi\'{c}, J.,  Vuleti\'{c}, V.,  Ye, J. \&  Zwierlein, M. Quantum simulators: Architectures and opportunities. 
\href{https://doi.org/10.1103/PRXQuantum.2.017003}{{\it PRX Quantum}~{\bf 2}, 017003 (2021)}.
%
\bibitem{Kitaev2003} 
Kitaev, A. Y. Fault-tolerant quantum computation by anyons. 
\href{https://doi.org/10.1016/S0003-4916(02)00018-0}{{\it Ann. Phys.}~{\bf 303}, 2 (2003)}.
%
\bibitem{Semeghini2021} 
Semeghini, G, Levine, H., Keesling, A., Ebadi, S., Wang, T. T., Bluvstein, D., Verresen, R., Pichler, H., Kalinowski, M., Samajdar, R., Omran, A., Sachdev, S., Vishwanath, A., Greiner, M., Vuleti\'{c}, V. \& Lukin, M. D. Probing topological spin liquids on a programmable quantum simulator. \href{https://doi.org/10.1126/science.abi8794}{{\it Science}~{\bf 374}, 1242 (2021)}.
%
\bibitem{Cao2018a}
Cao, Y., Fatemi, V., Demir, A., Fang, S., Tomarken, S. L., Luo, J. Y., Sanchez-Yamagishi, J. D., Watanabe, K., Taniguchi, T., Kaxiras, E., Ashoori, R. C. \& Jarillo-Herrero, P. Correlated insulator behaviour at half-filling in magic-angle graphene superlattices. \href{https://doi.org/10.1038/nature26154}{{\it Nature}~{\bf 556}, 80 (2018)}.
%
\bibitem{Cao2018b} 
Cao, Y., Fatemi, V., Fang, S., Watanabe, K., Taniguchi, T., Kaxiras, E. \& Jarillo-Herrero, P. Unconventional superconductivity in magic-angle graphene superlattices. 
\href{https://doi.org/10.1038/nature26160}{{\it Nature}~{\bf 556}, 43 (2018)}.
%
\bibitem{Andrei2021} 
Andrei, E. Y., Efetov, D. K., Jarillo-Herrero, P., MacDonald, A. H., Mak, K. F., Senthil, T., Tutuc, E., Yazdani, A. \& Young, A. F. The marvels of moir\'{e} materials. \href{https://doi.org/10.1038/s41578-021-00284-1}{{\it Nat. Rev. Mater.}~{\bf 6}, 201 (2021)}.
%
\bibitem{Bistritzer2011} 
Bistritzer, R. \& MacDonald, A. H. Moir\'{e} bands in twisted double-layer graphene. 
\href{https://doi.org/10.1073/pnas.1108174108}{{\it Proc. Natl. Acad. Sci. (USA)}~{\bf 108}, 12233 (2011)}.
%
\bibitem{Lu2019}
Lu, X., Stepanov, P., Yang, W., Xie, M., Aamir, M. A., Das, I., Urgell, C., Watanabe, K., Taniguchi, T., Zhang, G., Bachtold, A., MacDonald, A. H. \& Efetov, D. K. Superconductors, orbital magnets and correlated states in magic-angle bilayer graphene. \href{https://doi.org/10.1038/s41586-019-1695-0}{{\it Nature}~{\bf 574}, 653 (2019)}.
%
\bibitem{Park2021} 
Park, J. M., Cao, Y., Watanabe, K., Taniguchi, T. \& Jarillo-Herrero, P. Tunable strongly coupled superconductivity in magic-angle twisted trilayer graphene. \href{https://doi.org/10.1038/s41586-021-03192-0}{{\it Nature}~{\bf 590}, 249 (2021)}.
%
\bibitem{Zao2021} 
Hao, Z., Zimmerman, A. M., Ledwith, P., Khalaf, E., Najafabadi, D. H., Watanabe, K., Taniguchi, T., Vishwanath, A. \& Kim, P. Electric field tunable unconventional superconductivity in alternating twist magic-angle trilayer graphene. \href{https://doi.org/10.1126/science.abg0399
}{{\it Science}~{\bf 371}, 1133 (2021)}.
%
\bibitem{Cao2021} 
Cao, Y., Park, J. M., Watanabe, K., Taniguchi, T. \& Jarillo-Herrero, P. Pauli-limit violation and re-entrant superconductivity in moir\'{e} graphene. \href{https://doi.org/10.1038/s41586-021-03685-y}{{\it Nature}~{\bf 595}, 526 (2021)}.
%
\bibitem{Hesp2021}
Hesp, N. C. H., Torre, I., Rodan-Legrain, D., Novelli, P., Cao, Y., Carr, S., Fang, S., Stepanov, P., Barcons-Ruiz, D., Herzig-Sheinfux, H., Watanabe, K., Taniguchi, T., Efetov, D. K., Kaxiras, E, Jarillo-Herrero, P., Polini, M. \& Koppens, F. H. L. Observation of interband collective excitations in twisted bilayer graphene. \href{https://doi.org/10.1038/s41567-021-01327-8}{{\it Nat. Phys.}~{\bf 17}, 1162 (2021)}.
%
\bibitem{Ni2018} 
Ni, G. X., McLeod, A. S., Sun, Z., Wang, L., Xiong, L, Post, K. W., Sunku, S. S., Jiang, B.-Y., Hone, J., Dean, C. R., Fogler, M. M. \& Basov, D. N. Fundamental limits to graphene plasmonics. \href{https://doi.org/10.1038/s41586-018-0136-9}{{\it Nature}~{\bf 557}, 530 (2018)}.
%
\bibitem{Sun2020} 
Sun, Z., Fogler, M. M., Basov, D. N. \& Millis, A. J. Collective modes and terahertz near-field response of superconductors. \href{https://doi.org/10.1103/PhysRevResearch.2.023413}{Phys. Rev. Res.~{\bf 10}, 023413 (2020)}.
%
\bibitem{Kennes2021} 
Kennes, D. M., Claassen, M., Xian, L., Georges, A., Millis, A. J., Hone, J., Dean, C. R., Basov, D. N., Pasupathy, A. \& Rubio, A. Moir\'{e} heterostructures as a condensed-matter quantum simulator. \href{https://doi.org/10.1038/s41567-020-01154-3}{{\it Nat. Phys.}~{\bf 17}, 155 (2021)}.
%
\bibitem{Sharpe2019} 
Sharpe, A. L., Fox, E. J., Barnard, A. W., Finney, J., Watanabe, K., Taniguchi, T, Kastner, M. A. \& Goldhaber-Gordon, D. Emergent ferromagnetism near three-quarters filling in twisted bilayer graphene. \href{10.1126/science.aaw3780}{{\it Science}~{\bf 365}, 605 (2019)}.
%
\bibitem{Stepanov2020} 
Stepanov, P., Das, I., Lu, X., Fahimniya, A., Watanabe, K., Taniguchi, T., Koppens, F. H. L.,  Lischner, J., Levitov, L. \& Efetov, D. K., Untying the insulating and superconducting orders in magic-angle graphene. \href{https://doi.org/10.1038/s41586-020-2459-6}{{\it Nature}~{\bf 583}, 375 (2020)}.
%
\bibitem{RodanLegrain2021} 
Rodan-Legrain, D., Cao, Y., Park, J. M., de la Barrera, S. C., Randeria, M. T., Watanabe, K., Taniguchi, T. \& Jarillo-Herrero, P. Highly tunable junctions and nonlocal Josephson effect in magic angle graphene tunneling devices. \href{https://doi.org/10.1038/s41565-021-00894-4}{\href{https://doi.org/10.1038/s41565-021-00894-4}{Nat. Nanotech.}~{\bf 16}, 769 (2021)}.
%
\bibitem{deVries2021} 
de Vries, F. K., Portol\'{e}s, E., Zheng, G., Taniguchi, T., Watanabe, K., Ihn, T., Ensslin, K. \& Rickhaus, P. Gate-defined Josephson junctions in magic-angle twisted bilayer graphene. \href{https://doi.org/10.1038/s41565-021-00896-2}{{\it Nat. Nanotech.}~{\bf 16}, 760 (2021)}.
%
\bibitem{Deng2020} 
Deng, B., Ma, C., Wang, Q., Yuan, S., Watanabe, K., Taniguchi, T., Zhang, F. \& Xia, F. Strong mid-infrared photoresponse in small-twist-angle bilayer graphene. \href{https://doi.org/10.1038/s41566-020-0644-7}{{\it Nat. Photon.}~{\bf 14}, 549 (2020)}.
%
\bibitem{DiBattista2021} 
Di Battista, G., Seifert, P., Watanabe, K., Taniguchi, T., Fong, K. C., Principi, A. \& Efetov, D. K. Revealing the ultra-sensitive calorimetric properties of superconducting magic-angle twisted bilayer graphene. \href{https://arxiv.org/abs/2111.08735}{{\it arXiv:2111.08735} (2021)}.
%
\bibitem{Uri2020} 
Uri, A., Grover, S., Cao, Y., Crosse, J. A., Bagani, K., Rodan-Legrain, D., Myasoedov, Y., Watanabe, K., Taniguchi, T., Moon, P., Koshino, M., Jarillo-Herrero, P. \& Zeldov, E. Mapping the twist-angle disorder and Landau levels in magic-angle graphene, \href{https://doi.org/10.1038/s41586-020-2255-3}{{\it Nature}~{\bf 581}, 47 (2020)}.
%
\bibitem{Masubuchi2018} 
Masubuchi, S., Morimoto, M., Morikawa, S., Onodera, M., Asakawa, Y., Watanabe, K., Taniguchi, T. \& Machida, T. Autonomous robotic searching and assembly of two-dimensional crystals to build van der Waals superlattices. \href{https://doi.org/10.1038/s41467-018-03723-w}{{\it Nat. Commun.}~{\bf 9}, 1413 (2018)}.
%
\bibitem{Han2020} 
Han, B., Lin, Y., Yang, Y., Mao, N., Li, W., Wang, H., Yasuda, K., Wang, X., Fatemi, V., Zhou, L., Wang, J. I.J., Ma, Q., Cao, Y., Rodan-Legrain, D., Bie, Y. Q., Navarro-Moratalla, E., Klein, D., MacNeill, D., Wu, S., Kitadai, H., Ling, X., Jarillo-Herrero, P., Kong, J., Yin, J. \& Palacios, T. Deep learning enabled fast optical characterization of two-dimensional materials. \href{https://doi.org/10.1002/adma.202000953}{{\it Adv. Mater.}~{\bf 32}, 2000953 (2020)}.
%
\bibitem{Drienovsky2018} 
Drienovsky, M., Joachimsmeyer, J., Sandner, A., Liu, M.-H., Taniguchi, T., Watanabe, K., Richter, K., Weiss, D. \& Eroms, J. Commensurability oscillations in one-dimensional graphene superlattices. \href{https://doi.org/10.1103/PhysRevLett.121.026806}{{\it Phys. Rev. Lett.}~{\bf 121}, 026806 (2018)}.
%
\bibitem{Song2018} 
Song, J. C. W. \& Gabor, N. M. Electron quantum metamaterials in van der Waals heterostructures. \href{https://doi.org/10.1038/s41565-018-0294-9}{{\it Nat. Nanotech.}~{\bf 13}, 986 (2018)}.
%
\bibitem{Forsythe2018} 
Forsythe, C., Zhou, X., Watanabe, K., Taniguchi, T., Pasupathy, A, Moon, P., Koshino, M., Kim, P. \& Dean, C. R. Band structure engineering of 2D materials using patterned dielectric superlattices. \href{https://doi.org/10.1038/s41565-018-0138-7}{{\it Nat. Nanotech.}~{\bf 13}, 566 (2018)}.
%
\bibitem{Huber2020} 
Huber, R., Liu, M.-H., Chen, S.-C., Drienovsky, M., Sandner, A., Watanabe, K., Taniguchi, T., Richter, K., Weiss, D. \& Eroms, J. Gate-tunable two-dimensional superlattices in graphene. \href{https://doi.org/10.1021/acs.nanolett.0c03021}{{\it Nano Lett.}~{\bf 20}, 8046 (2020)}.
%
\bibitem{Li2021} 
Li, Y., Dietrich, S., Forsythe, C., Taniguchi, T., Watanabe, K., Moon, P. \& Dean, C. R. Anisotropic band flattening in graphene with one-dimensional superlattices. \href{https://doi.org/10.1038/s41565-021-00849-9}{{\it Nat. Nanotech.}~{\bf 16}, 525 (2021)}.
%
\bibitem{Shi2019} 
Shi, L.-k., Ma, J. \& Song, J. C. W. Gate-tunable flat bands in van der Waals patterned dielectric superlattices. \href{https://doi.org/10.1088/2053-1583/ab59a8}{{\it 2D Mater.}~{\bf 7}, 015028 (2019)}.
%
\bibitem{Ho2018} 
Ho, A., McClean, J. \& Ong, S. P. The promise and challenges of quantum computing for energy storage. \href{https://doi.org/10.1016/j.joule.2018.04.021}{{\it Joule}~{\bf 2}, 810 (2018)}.
%
\bibitem{Ajagekar2019} 
Ajagekar, A. \& You, F. Quantum computing for energy systems optimization: Challenges and opportunities. \href{https://doi.org/10.1016/j.energy.2019.04.186}{{\it Energy}~{\bf 179}, 76 (2019)}.
%
\bibitem{Scully2011} 
Scully, M. O., Chapin, K. R., Dorfman, K. E., Kim, M. B., and Svidzinsky, A. Quantum heat engine power can be increased by noise-induced coherence. \href{https://doi.org/10.1073/pnas.1110234108}{{\it Proc. Natl. Acad. Sci. (USA)}~{\bf 108}, 15097 (2011)}.
%
\bibitem{Bittner2014} 
Bittner, E. R. \& Silva, C. Noise-induced quantum coherence drives photo-carrier generation dynamics at polymeric semiconductor heterojunctions. \href{https://doi.org/10.1038/ncomms4119}{{\it Nat. Commun.}~{\bf 5}, 3119 (2014)}.
%
\bibitem{Chin2013}
Chin, A. W., Prior, J., Rosenbach, R., Caycedo-Soler, F., Huelga, S. F. \& Plenio, M. B. The role of non-equilibrium vibrational structures in electronic coherence and recoherence in pigment-protein complexes. \href{https://doi.org/10.1038/nphys2515}{{\it Nat. Phys.}~{\bf 9}, 113 (2013)}.
%
\bibitem{Dorfman2013}
Dorfman, K. E., Voronine, D. V., Mukamel, S. \& Scully. M. O. Photosynthetic reaction center as a quantum heat engine. \href{https://doi.org/10.1073/pnas.1212666110}{{\it Proc. Natl. Acad. Sci. (USA)}~{\bf 110}, 2746 (2013)}.
%
\bibitem{Zhong2016} 
Zhong, X., Chervy, T., Wang, S., George, J., Thomas, A., Hutchison, J. A., Devaux, E., Genet, C. \& Ebbesen, T. W. Non-radiative energy transfer mediated by hybrid light-matter states. \href{https://doi.org/10.1002/anie.201600428}{{\it Angew. Chem. Int. Ed.}~{\bf 55}, 6202 (2016)}.
%
\bibitem{Zhong2017} 
Zhong, X., Chervy, T., Zhang, L., Thomas, A., George, J., Genet, C., Hutchison, J. A. \& Ebbesen, T. W. Energy transfer between spatially separated entangled molecules. \href{https://doi.org/10.1002/anie.201703539}{{\it Angew. Chem. Int. Ed.}~{\bf 56}, 9034 (2017)}.
%
\bibitem{Rozenman2018} 
Rozenman, G. G., Akulov, K., Golombek, A. \& Schwartz, T. Long-range transport of organic exciton-polaritons revealed by ultrafast microscopy. \href{https://doi.org/10.1021/acsphotonics.7b01332}{{\it ACS Photon.}~{\bf 5}, 105 (2018)}.
%
\bibitem{SaezBlazquez2018}
S\'{a}ez-Bl\'{a}zquez R., Feist J., Fern\'{a}ndez Dom\'{i}nguez A. I., Garc\'{i}a-Vidal F. J. Organic polaritons enable local vibrations to drive long-range energy transfer. \href{https://doi.org/10.1103/PhysRevB.97.241407}{{\it Phys. Rev. B}~{\bf 97}, 241407(R) (2018)}.
%
\bibitem{Rice2021} 
Rice, J. E., Gujarati, T. P., Motta, M. , Takeshita, T. Y., Lee, E., Latone, J. A. \& Garcia, J. M. Quantum computation of dominant products in lithium-sulfur batteries. \href{https://doi.org/10.1063/5.0044068}{{\it J. Chem. Phys.}~{\bf 154}, 134115 (2021)}.
%
\bibitem{Gao2021} 
Gao, Q., Nakamura, H., Gujarati, T. P., Jones, G. O., Rice, J. E., Wood, S. P., Pistoia, M., Garcia, J. M. \& Yamamoto, N. Computational investigations of the lithium superoxide dimer rearrangement on noisy quantum devices. \href{https://doi.org/10.1021/acs.jpca.0c09530}{{\it J. Phys. Chem. A}~{\bf 125}, 1827 (2021)}.
%
\bibitem{Benenti2017} 
Benenti, G., Casati, G., Saito, K. \& Whitney, R. S. Fundamental aspects of steady-state conversion of heat to work at the nanoscale. \href{https://doi.org/10.1016/j.physrep.2017.05.008}{{\it Phys. Rep.}~{\bf 694}, 1 (2017)}.
%
\bibitem{Pekola2021} 
Pekola, J. P. \& Karimi, B. Quantum heat transport in condensed matter systems. \href{https://doi.org/10.1103/RevModPhys.93.041001}{{\it Rev. Mod. Phys.}~{\bf 93}, 041001 (2021)}.
%
\bibitem{Josefsson2018} 
Josefsson, M., Svilans, A., Burke, A. M., Hoffmann, E. A., Fahlvik, S., Thelander, C., Leijnse, M. \& Linke, H. A quantum-dot heat engine operating close to the thermodynamic efficiency limits. \href{https://doi.org/10.1038/s41565-018-0200-5}{{\it Nat. Nanotech.}~{\bf 13}, 920 (2018)}.
%
\bibitem{vonLindenfels2019} 
von Lindenfels, D., Gr\"{a}b, O., Schmiegelow, C. T., Kaushal, V., Schulz, J., Mitchison, M. T., Goold, J., Schmidt-Kaler, F. \& Poschinger, U. G. Spin heat engine coupled to harmonic-oscillator flywheel. \href{https://doi.org/10.1103/PhysRevLett.123.080602}{{\it Phys. Rev. Lett.}~{\bf 123}, 080602 (2019)}.
%
\bibitem{Klatzow2019} 
Klatzow, J., Becker, J. N., Ledingham, P. M., Weinzetl, C., Kaczmarek, K. T., Saunders, D. J., Nunn, J., Walmsley, I. A., Uzdin, R. \& Poem, E. Experimental demonstration of quantum effects in the operation of microscopic heat engines. \href{https://doi.org/10.1103/PhysRevLett.122.110601}{{\it Phys. Rev. Lett.}~{\bf 122}, 110601 (2019)}.
%
\bibitem{Prete2019} 
Prete, D., Erdman, P. A., Demontis, V., Zannier, V., Ercolani, D., Sorba, L., Beltram, F., Rossella, F., Taddei, F. \& Roddaro, S. Thermoelectric conversion at 30 K in InAs/InP nanowire quantum dots. \href{https://doi.org/10.1021/acs.nanolett.9b00276}{{\it Nano Lett.}~{\bf 19}, 3033 (2019)}.
%
\bibitem{Ono2020} 
Ono, K., Shevchenko, S. N., Mori, T., Moriyama, S. \& Nori, F. Analog of a quantum heat engine using a single-spin qubit. \href{https://doi.org/10.1103/PhysRevLett.125.166802}{{\it Phys. Rev. Lett.}~{\bf 125}, 166802 (2020)}.
%
\bibitem{Guthrie2021} 
Guthrie, A., Satrya, C. D., Chang, Y.-C., Menczel, P., Nori, F. \& Pekola, J. P. 
A Cooper-pair box architecture for cyclic quantum heat engines, \href{https://arxiv.org/abs/2109.03023}{{\it arXiv:2109.03023} (2021)}.
%
\bibitem{Scharf2020} 
Scharf, B., Braggio, A., Strambini, E, Giazotto, F. \& Hankiewicz, E. M. Topological Josephson heat engine. \href{https://doi.org/10.1038/s42005-020-00463-6}{{\it Commun. Phys.}~{\bf 3}, 198 (2020)}.
%
\bibitem{Marchegiani2020} 
Marchegiani, G., Braggio, A. \& Giazotto, F. Superconducting nonlinear thermoelectric heat engine. \href{https://doi.org/10.1103/PhysRevB.101.214509}{{\it Phys. Rev. B}~{\bf 101}, 214509 (2020)}.
%
\bibitem{Alicki2013} 
Alicki, R. \& Fannes, M. Entanglement boost for extractable work from ensembles of quantum batteries. \href{https://doi.org/10.1103/PhysRevE.87.042123}{{\it Phys. Rev. E}~{\bf 87}, 042123 (2013)}.
%
\bibitem{Hovhannisyan2013} 
Hovhannisyan, K. V., Perarnau-Llobet, M., Huber, M. \& Ac\'{i}n, A. Entanglement generation is not necessary for optimal work extraction. \href{https://doi.org/10.1103/PhysRevLett.111.240401}{{\it Phys. Rev. Lett.}~{\bf 111}, 240401 (2013)}.
%
\bibitem{Binder2015} 
Binder, F. C., Vinjanampathy, S., Modi, K.  \& Goold, J. Quantacell: powerful charging of quantum batteries. \href{https://doi.org/10.1088/1367-2630/17/7/075015}{{\it New J. Phys.}~{\bf 17}, 075015 (2015)}.
%
\bibitem{Campaioli2017} 
Campaioli, F., Pollock, F. A., Binder, F. C., C\'{e}leri, L., Goold, J., Vinjanampathy, S. \& Modi, K. Enhancing the charging power of quantum batteries. \href{https://doi.org/10.1103/PhysRevLett.118.150601}{{\it Phys. Rev. Lett.}~{\bf 118}, 150601 (2017)}.
%
\bibitem{Campaioli2018} 
For a recent review on quantum batteries see e.g. Campaioli, F., Pollock, F. A. \& Vinjanampathy, S. in ``\href{https://doi.org/10.1007/978-3-319-99046-0}{Thermodynamics in the Quantum Regime}'', Binder, F., Correa, L. A., Gogolin, C., Anders, J. \& Adesso, G. (eds.), p. 207-225, (Springer, 2018).
%
\bibitem{Ferraro2018} 
Ferraro, D., Campisi, M., Andolina, G. M., Pellegrini, V. \& Polini, M. High-power collective charging of a solid-state quantum battery. \href{https://doi.org/10.1103/PhysRevLett.120.117702}{{\it Phys. Rev. Lett.}~{\bf 120}, 117702 (2018)}.
%
\bibitem{Andolina2019} 
Andolina, G. M., Keck, M., Mari, A., Campisi, M., Giovannetti, V. \& Polini, M. Extractable work, the role of correlations, and asymptotic freedom in quantum batteries. \href{https://doi.org/10.1103/PhysRevLett.122.047702}{{\it Phys. Rev. Lett.}~{\bf 122}, 047702 (2019)}.%
%
\bibitem{Sergi2020}
Juli\`{a}-Farr\'{e}, S., Salamon, T., Riera, A., Bera, M. N. \& Lewenstein, M. Bounds on the capacity and power of quantum batteries. \href{https://doi.org/10.1103/PhysRevResearch.2.023113}{{\it Phys. Rev. Research}~{\bf 2}, 023113 (2020)}.
%
\bibitem{Rossini2020} 
Rossini, D., Andolina, G. M., Rosa, D., Carrega, M. \& Polini, M. Quantum advantage in the charging process of Sachdev-Ye-Kitaev batteries. \href{https://doi.org/10.1103/PhysRevLett.125.236402}{{\it Phys. Rev. Lett.}~{\bf 125}, 236402 (2020)}.
%
\bibitem{Quach2020} 
Quach, J. Q., McGhee, K. E., Ganzer, L., Rouse, D. M., Lovett, B. W., Gauger, E. M., Keeling, J., Cerullo, G., Lidzey, D. G. \& Virgili, T. Superabsorption in an organic microcavity: Toward a quantum battery. \href{https://doi.org/10.1126/sciadv.abk3160}{Sci. Adv.~{\bf 8}, eabk3160 (2022)}.
%
\bibitem{Hu2021} 
Hu, C.-K., Qiu, J., Souza, P. J. P., Yuan, J., Zhou, Y., Zhang, L., Chu, J., Pan, X., Hu, L., Li, J., Xu, Y., Zhong, Y., Liu, S., Yan, F., Tan, D., Bachelard, R., Villas-Boas, C.  J., Santos, A. C. \& Yu, D. Optimal charging of a superconducting quantum battery. \href{https://arxiv.org/abs/2108.04298}{{\it arXiv:2108.04298} (2021)}.
%
\bibitem{Ferraro2019} 
Ferraro, D., Andolina, G. M., Campisi, M., Pellegrini, V. \& Polini, M. Quantum supercapacitors. \href{https://doi.org/10.1103/PhysRevB.100.075433}{{\it  Phys. Rev. B}~{\bf 100}, 075433 (2019)}.
%
\bibitem{Tirone2021} 
Tirone, S., Salvia, R. \& Giovannetti, V. Quantum energy lines and the optimal output ergotropy problem. \href{https://doi.org/10.1103/PhysRevLett.127.210601}{{\it Phys. Rev. Lett.}~{\bf 127}, 210601 (2021)}.
%
\bibitem{GarciaVidal2021} 
Garcia-Vidal, F. J., Ciuti, C. \& Ebbesen, T. W. Manipulating matter by strong coupling to vacuum fields. \href{https://doi.org/10.1126/science.abd0336}{{\it Science}~{\bf 373}, eabd0336 (2021)}.
%
\bibitem{Hubener2021} 
H\"{u}bener, H., De Giovannini, U., Sch\"{a}fer, C., Andberger, J., Ruggenthaler, M., Faist, J. \& Rubio, A. Engineering quantum materials with chiral optical cavities. \href{https://doi.org/10.1038/s41563-020-00801-7}{{\it Nat. Mater.}~{\bf 20}, 438 (2021)}.
%
\bibitem{Thomas2019} 
Thomas, A., Lethuillier-Karl, L., Nagarajan, K., Vergauwe, R. M. A., George, J., Chervy, T., Shalabney, A., Devaux, E., Genet, C., Moran, J. \& Ebbesen, T. W. Tilting a ground-state reactivity landscape by vibrational strong coupling. \href{https://doi.org/10.1126/science.aau7742}{{\it Science}~{\bf 363}, 615 (2019)}.
%
\bibitem{Thomas2021}
Thomas, A., Devaux, E., Nagarajan, K., Rogez, G., Seidel, M., Richard, F., Genet, C., Drillon, M. \& Ebbesen, T. W. Large enhancement of ferromagnetism under a collective strong coupling of YBCO nanoparticles. \href{https://doi.org/10.1021/acs.nanolett.1c00973}{{\it Nano Lett.}~{\bf 21}, 4365 (2021)}.
%
\bibitem{Paravicini2019} 
Paravicini-Bagliani, G. L., Appugliese, F., Richter, E., Valmorra, F., Keller, J., Beck, M., Bartolo, N., R\"{o}ssler, C., Ihn, T., Ensslin, K., Ciuti, C., Scalari, G. \& Faist, J. Magneto-transport controlled by Landau polariton states. \href{https://doi.org/10.1038/s41567-018-0346-y}{{\it Nat. Phys.}~{\bf 15}, 186 (2019)}.
%
\bibitem{Ikonen2017} 
Ikonen, J., Salmilehto, J. \& M\"{o}tt\"{o}nen, M. Energy-efficient quantum computing. \href{https://doi.org/10.1038/s41534-017-0015-5}{{\it npj Quantum Inf.}~{\bf 3}, 17 (2017)}.
%
\bibitem{Chiribella2021} 
Chiribella, G., Yang, Y. \& Renner, R. Fundamental energy requirement of reversible quantum operations. \href{https://doi.org/10.1103/PhysRevX.11.021014}{{\it Phys. Rev. X}~{\bf 11}, 021014 (2021)}.
%
\bibitem{Fellous2021} 
Fellous-Asiani, M., Chai, J. H., Whitney, R. S., Auff\`{e}ves, A. \& Ng, H. K. Limitations in quantum computing from resource constraints. \href{https://doi.org/10.1103/PRXQuantum.2.040335}{{\it PRX Quantum}~{\bf 2}, 040335 (2021)}.
%
\end{thebibliography}
\end{document}